\documentclass[a4paper,11pt]{article}

\usepackage{jheppub1}
\usepackage{array}
\usepackage{bbold}
\usepackage{multirow}
\usepackage{amsmath}
\usepackage{makecell}
\usepackage{braket}
\usepackage{enumerate}
  {\left.\begin{aligned}}
  {\end{aligned}\right\rbrace}

\definecolor{green}{rgb}{0.1,0.8,0.2}

\usepackage{cancel}
\usepackage{amsmath,amssymb} 
\usepackage{graphicx} 
\usepackage[dvipsnames,x11names]{xcolor} 

\usepackage{hyperref}
\usepackage{cleveref}
\usepackage{float}
\usepackage{subcaption}
\usepackage{tikz, pgfplots}
\usepackage{caption}
\usetikzlibrary{snakes}

\makeatletter
\newcommand{\footnoteref}[1]{\protected@xdef\@thefnmark{\ref{#1}}\@footnotemark}
\makeatother

\begin{document}

\preprint{}
\title{Moving Mirrors, OTOCs and Scrambling}
\author[a]{Parthajit Biswas}
    \affiliation[a]{Department of Physics, Ramakrishna Mission Vivekananda Educational and Research Institute, Belur Math, Howrah 711202, India.}
\author[a]{, Bobby Ezhuthachan} 
\author[b,c,d]{, Arnab Kundu}
\affiliation[b]{Saha Institute of Nuclear Physics, 1/AF Bidhannagar, Kolkata 700064, India.}
\affiliation[c]{Homi Bhabha National Institute, Training School Complex, Anushaktinagar, Mumbai 400094, India.}
\affiliation[d]{
Institut f\"{u}r Theoretische Physik und Astrophysik, Julius-Maximilians-Universit\"{a}t W\"{u}rzburg, Am Hubland, 97074 W\"{u}rzburg, Germany.}
\author[a]{, Baishali Roy}
\emailAdd{parthajitbiswas8@gmail.com, bobby.phy@gm.rkmvu.ac.in, arnab.kundu@saha.ac.in, baishali.roy025@gm.rkmvu.ac.in}

\abstract{We explore the physics of scrambling in the moving mirror models, in which a two-dimensional CFT is subjected to a time-dependent boundary condition. It is well-known that by choosing an appropriate mirror profile, one can model quantum aspects of black holes in two-dimensions, ranging from Hawking radiation in an eternal black hole (for an ``escaping mirror") to the recent realization of Page curve in evaporating black holes (for a ``kink mirror"). We explore a class of OTOCs in the presence of such a boundary and explicitly demonstrate the following primary aspects: First, we show that the dynamical CFT data directly affect an OTOC and maximally chaotic scrambling occurs for the escaping mirror for a large-$c$ CFT with identity block dominance. We further show that the exponential growth of OTOC associated with the physics of scrambling yields a power-law growth in the model for evaporating black holes which demonstrates a unitary dynamics in terms of a Page curve. We also demonstrate that, by tuning a parameter, one can naturally interpolate between an exponential growth associated to scrambling and a power-law growth in unitary dynamics. Our work explicitly exhibits the role of higher-point functions in CFT dynamics as well as the distinction between scrambling and Page curve. We also discuss several future possibilities based on this class of models.}

\maketitle

%%%%%%%%%%%%%%%%%%
%\section{Introduction}
%%%%%%%%%%%%%%%%%%

%1. Broad remarks on the role of boundary conditions on chaos , etc.

%2. In general a difficult problem. Boundary in 2d CFT is also related to quench dynamics. Discuss how these are useful in CFT, etc.

%3. Mention the role of moving mirror models in a general context.

%4. The physics of scrambling: loss and gain of the unitary information.

%5. Discuss our results.

%6. More remarks and discussions on the results. 

%7. etc

%In classical dynamics, we can make a non-chaotic system chaotic by changing boundary condition. In this work, we want to investigate the same for quantum system with time dependent boundary. We consider a CFT with time dependent boundary. Following \cite{Das:2019tga}, we compute four-point out-of-time-ordered correlator (OTOC) for zero temperature and finite temperature field theory. For, zero temperature case, we see that the four-point OTOC shows exponential decay with Lyapunov index determined by the form of the time dependent boundary. Which suggests that in zero temperature field theory, time dependent boundary condition induces effective temperature to the system. For finite temperature case, OTOC again shows exponential decay with Lyapunov index which is different from that of the thermal case without the time dependent boundary.

%%%%%%%%%%%%%%%%%%%%%%%%%%%%%%
\section{Introduction}
%%%%%%%%%%%%%%%%%%%%%%%%%%%%%%

It is well-known that a classical integrable dynamics can exhibit chaotic features depending on the boundary conditions. The prototypical example of this is the billiard ball problem, where a free particle can become ergodic depending on the shape of the stadium and therefore the boundary conditions imposed on the dynamics. The closest quantum analogue of diverging nearby trajectories is given by measuring expectation values of non-commuting observables in quantum mechanics. Consider two observers, $A$ and $B$ who share a common physical state $| \psi \rangle$. If observer $A$ wants to communicate one bit $a$ to observer $B$, she can apply a unitary operation to the quantum state $| \psi \rangle$. Subsequently, the resulting state evolves in time, and observer $B$ makes a measurement on the evolved state. The probability of measuring a certain bit-information now can be written in terms of an expectation value of the commutator of the two operators, in the state $| \psi \rangle$.\footnote{more precisely, the probability is given by \cite{Xu:2022vko}
\begin{eqnarray}
P\left(B|A \right) = \left \langle \psi \right | U_a^\dagger [P_b(t), U_a] \left |\psi \right \rangle \ ,
\end{eqnarray}
where $P_b$ is the measurement operator yielding the result $b$, as measured by observer $B$ and $$P_b(t) \equiv e^{i H t} P_b e^{- i H t}\,,$$ while $U_{a}$ is a unitary operation performed on the state by observer $A.$} The expectation value of the absolute value of commutator-squared of any two operators can be written in terms of usual time-order correlators and out-of-time-order correlators, or OTOCs in brief. 

The dynamical behaviour of the OTOC quantitatively defines scrambling, which relates to the transfer of local information to non-local degrees of freedom, and its associated time-scale, see {\it e.g.} \cite{Yoshida:2017non}.\footnote{Note that, as is discussed in detail in \cite{Yoshida:2017non}, the connection of OTOC to perfect scrambling becomes precise when a Haar-random averaging is carried out over the unitary evolution of operators. Such averaged OTOCs are then related to R\'{e}nyi $2$-mutual information. The decay of the corresponding OTOC is therefore related to a faithful recovery of the state.}  Relatedly, scrambling of a localized information across a given system is connected to thermalization and a notion of ``early time chaos". It is further conjectured that black holes are the fastest scramblers in Nature \cite{Sekino:2008he}, in which the scrambling time scales logarithmically in the number of degrees of freedom of the system: $t_{\rm scramble} \sim \log N$; whereas for typical quantum mechanical and quantum field theoretic systems, this scales as a power-law: $t_{\rm scramble} \sim N^{\alpha}$, where $\alpha$ is a real number.\footnote{Note that, exponential growth is still possible to obtain in weakly-coupled QFTs, provided the contribution from an infinite diagrams can be resummed. These infinite diagrams can be obtained as a result of a large-$N$ expansion, see {\it e.g.}~\cite{Stanford:2015owe}. The exponent of the exponential growth is nonetheless well below the chaos bound of \cite{Maldacena:2015waa}. In this article, we will not address the possibility of a resummation of perturbative contributions that yield a power-law in time. It is an intriguing aspect to which we hope to return in future.} 

%However no simple and sharp statement exists in terms of exponential growth of such correlators and their dependence on the boundary conditions of the system. 

Note that, while the notion of quantum chaos is multi-faceted and non-unique, OTOCs have recently been studied extensively in the literature from a wide perspective and within a very large class of models. While early studies of OTOCs include \cite{Larkin:1964wok}, more recently, an interesting bound, known as the MSS bound \cite{Maldacena:2015waa}, have been derived on the growth of such correlators, based on unitarity and causality. Given this, an already interesting aspect is to explore the role of boundary conditions on such correlation functions and their respective bounds. In a general QFT, however, this is a technically difficult task for two reasons: First, it is difficult to calculate explicitly the Lyapunov growth of an OTOC in a perturbative framework of QFT. Typically, one needs a large parameter (such as $N$ of a large-$N$ theory) in which an expansion can be carried out and a resummation of such a perturbation theory usually contains the information of chaotic features. Note that, one needs to sum over an infinite class of diagrams and therefore the OTOC growth is insensitive to a standard perturbative weak-coupling calculation in QFT. See {\it e.g.}~\cite{Stanford:2015owe, Chowdhury:2017jzb}.\footnote{This aspect is somehow similar to computing non-trivial transport coefficients in perturbative QFT, {\it e.g.}~shear viscosity.}

Conformal field theories (CFTs) on the other hand provide us with an ideal framework for addressing this class of questions. While OTOCs are inherently four-point correlation functions where the time-ordering of the operators need to have a specific order and therefore are not completely fixed by the symmetries of the CFT, it is also encouraging that dynamical CFT data ({\it i.e.}~OPE data) are inherently encoded in a CFT OTOC. In fact, OTOCs can be used to diagnose and distinguish between the dynamics of an integrable CFT vs a chaotic one \cite{Das:2021qsd, Das:2022jrr} --- a feature that cannot be captured by a two-point or a three-point function which are universally fixed.

Furthermore, introducing a boundary in a two-dimensional CFT has a particularly rich set of possibilities: such boundaries can be defects, interfaces, or simply boundaries. Typically, these are distinguished by the boundary condition on the energy-momentum tensor of the corresponding CFT, or, more generally, on the Virasoro generators, at the location of the boundary. Consequently, several physical phenomena can be addressed within the framework of BCFT, including but not limited to critical quenches \cite{Calabrese:2006rx, Calabrese:2007mtj}, universal reflection and transmission coefficients through an interface \cite{Bachas:2020yxv, Karch:2023evr, Gillioz:2020mdd}, {\it etc}. Most of these examples are based on spatial boundaries (for interface CFTs) or boundaries in the Euclidean time ({\it e.g.}~for critical quench dynamics).

The class of models, known as the moving mirror models, on the other hand are based on boundaries that are dynamically evolving. The framework is inherently Lorentzian in which the mirrors are set on a timelike trajectory.\footnote{Technically, there is no problem in considering a spacelike or a null mirror trajectory, however, we do not consider these cases in the present work. See {\it e.g.}~\cite{Akal:2022qei} for more discussion on such trajectories.} These models have widely explored in the context of black hole Hawking radiation: early works in \cite{Davies:1977yv, PhysRevD.25.2569, PhysRevD.101.025012}, on Bogoliubov coefficients in \cite{PhysRevD.36.2327}, on later refinements and generalizations in \cite{Wilczek:1993jn}, on a stochastic perspective in \cite{Raval:1996vt}, on partner particles in \cite{PhysRevD.91.124060}, on entropy and information loss in \cite{Chen:2017lum, Reyes:2021npy, Akal:2020twv, Akal:2021foz}. On Casimir effect: on radiation reaction in \cite{Jaekel:1997hr}, on temperature corrections in \cite{Plunien:1999ba}, on time-dependence in \cite{PhysRevD.100.065022}. In cosmological expansion: On cosmological radiation in \cite{Casadio:2002dj}, on de-Sitter moving mirror cosmology in \cite{Good:2020byh}, production of entanglement entropy from moving mirror in \cite{Cotler:2022weg}; entanglement extraction in \cite{Cong:2018vqx, Cong:2020nec}, on moving interfaces in \cite{Biswas:2024xut} and many such. An extensive literature on the dual holographic construction of the moving mirror models also exists already \cite{Hirata:2008ka, Akal:2020twv, Akal:2021foz, Akal:2022qei}. In brief, it is fair to state that the moving mirror models are relevant in a broad class of interesting physical situations, and therefore understanding the physics of scrambling of this class of models is certainly a well-motivated and topical question.

In this article, we undertake this task focusing on instructive examples. Since the mirrors can be thought of as boundaries, we will focus on a specific class of three-point functions in the BCFT (referred to as the b-OTOC), which has been introduced in \cite{Das:2019tga}. While in a CFT without boundaries, three point functions are completely fixed by conformal symmetries up to a constant, in the presence of boundaries, this is not so. As was shown in \cite{Das:2019tga}, in large-c CFTs, the b-OTOCs show the characteristic Lyapunov growth in time, and thus provide a diagnostic for early-time chaos for these systems in the presence of a boundary. Mathematically, the reason why the b-OTOC works as a diagnostic for early-time chaos, despite being a three-point function is that in the presence of a boundary these three-point functions behave as a  holomorphic four-point functions (using the doubling trick), under conformal transformations. In particular, the ``doubled" four-point function so obtained, can be shown to be a OTOC. The Lyapunov behaviour then follows, from standard assumptions of vacuum block dominance.  

Using this as a  diagnostic, we demonstrate, in keeping with previous qualitative results in \cite{Das:2021qsd, Das:2022jrr}, that dynamical data of the CFT does demarcate between the behaviour of its corresponding OTOCs. For example, we explicitly demonstrate the following: In the ``escaping mirror" model, which captures features of Hawking radiation from an eternal black hole, a large-$c$ CFT displays an exponentially growing OTOC with a maximal Lyapunov exponent. However, a free boson or an Ising CFT does not exhibit any such growth. Thus, while the two-point function of all two-dimensional CFTs displays the same thermal feature in the escaping mirror model, there is a clear distinction between them at the level of the OTOC.

Furthermore, this result also suggests two intriguing complementary possibilities related to the quantum features of a black hole: (i) Fast scramblers such as black holes are not able to scramble strongly integrable dynamics of free bosons or the Ising CFT. There is a clear competition between the chaotic features of a black hole and the integrable structure of such examples. Note that, usually, it is believed that a black hole will act as a fast scrambler for any degree of freedom. However, strongly constrained systems ({\it e.g.}~integrable systems with an infinite number of conserved charges) may not scramble even in the presence of a black hole.\footnote{This feature has a qualitative resemblance to the ``scar-states" in an otherwise strongly thermalizing system. It will be very interesting to explore a more precise description of this connection.} (ii) Alternatively, the moving mirrors are unable to capture finer dynamical aspects of quantum black holes. While possibility (i) will have important implications in sharpening the scrambling features of a black hole, possibility (ii) is strongly suggestive of a limitation of the moving mirror models. At this point, however, we do not have an argument in favour of either of the possibilities above and we hope to address this in the near future.

Furthermore, in the ``kink" mirror model, we are able to explicitly demonstrate that the exponential growth of OTOC cannot set in. This model captures the dynamics of an evaporating black hole and it has been previously shown that the kink mirror model reproduces a unitary Page curve for the evolution of the fine-grained entropy \cite{Akal:2020twv}. It is thus expected that information recovery in this model should somehow be explicit. It is possible to obtain the ``escaping mirror" profile as a limit of the ``kink mirror", since the eternal black hole can also be obtained from the evaporating black hole in the limit when evaporation disappears. The mirror model captures a simple realization of this limit, in terms of one free parameter denoted by $u_0$. The escaping mirror is obtained in the limit $u_0 \to \infty$. We have explicitly demonstrated that in the escaping mirror limit the exponential growth of OTOC gets squeezed and eventually yields for a power-law growth for the kink mirror.

There is a simple picture that emerges from these models. In the eternal black hole, when no information recovery is possible by definition, we only observe the physics of information scrambling, {\it i.e.}~an exponential growth of the OTOC for a large-$c$ CFT. In other words, when the Page time, $t_{\rm Page}$, has been pushed to infinity, the scrambling time-scale, $t_{\rm scramble}$, is clearly visible\footnote{Evidently, a more precise statement is that scrambling is visible in the limit $t_{\rm Page} \gg t_{\rm scramble}$.} and $t_{\rm scramble} \sim \log (c)$. However, when evaporation of the black hole is introduced into the system, the time-dynamics of OTOCs are power-laws and therefore no natural scale can be associated with a fast-scrambling.\footnote{in other words, they are slow scramblers.} Furthermore, we also observe that in the $t_{\rm Page} \gg t_{\rm scramble}$ limit, the exponential growth is recovered; however, as the system is observed at a time-scale $t_{\rm observation} \sim t_{\rm Page}$, there is only a power law behaviour. %{\bf maybe we need to make it more manifest in the main draft?}

There is an intriguing parallel of the above observation to the scrambling and exponential behaviour of OTOCs in a weakly coupled QFT. In a weakly coupled perturbative framework, keeping only a few terms in the Feynman perturbation series generally do not reveal any exponentially growing piece in the OTOC. 
To observe the exponential behaviour, one needs to keep the contribution from several orders of the coupling constant \cite{Stanford:2015owe}. The parameter $u_0$ above is similar to the coupling constant in QFT:  When $u_0$ is large, {\it i.e.} the coupling is strong, we observe a fast scrambling behaviour. On the other hand, when $u_0 \sim {\cal O}(1)$, the scrambling becomes slow or is absent from the dynamics. %{\bf discuss about slow and fast scrambling a bit also.}
Thus, we have a simple model that can interpolate between a fast scrambling and a slow scrambling behaviour. 

%{\bf more discussions}

This article is divided into the following sections. After a brief review of the moving mirror setup in section \ref{2}, we present the main computation of the b-OTOC in the escaping mirror and the kink mirror profiles in section \ref{subsec:escaping} and \ref{subsec:kink} respectively.  We show that for the former mirror profile, a large-c CFT shows an exponential growth in the b-OTOC, while it shows a power law behaviour in the case of a kink profile. As we further show in section \ref{subsec:escaping}, this exponential growth in the escaping mirror profile, is not universal, in that, a similar computation in the Ising model and an orbifold CFT model, does not show an exponential growth. This is in contrast to the two-point function computation in these models, presented in section \ref{2pt} which is universal, showing a thermal behaviour for any CFTs in the escaping mirror background. In section \ref{subsec:escapingfromkink}, we further show, that by taking a particular limit of a parameter in the kink profile, one can reproduce the exponential behaviour in the b-OTOC. In section \ref{gem}, we discuss another generalization/deformation of the escaping mirror profile, where once again, the deformation parameter helps to interpolate between a scrambling and non-scrambling dynamics. We end with a summary of our results and a discussion on open issues and future directions to explore in section \ref{discussion}. Some details of the computations of the main draft are presented in appendix \ref{app:analytic}, while in appendix \ref{app: other} we generalize the b-OTOC computations of the main draft to other mirror profiles.

%%%%%%%%%%%%%%%%%%%%%%%%%%%%%%
\section{Moving mirror: preliminary remarks}\label{2}
%%%%%%%%%%%%%%%%%%%%%%%%%%%%%%

We begin with a brief review of the framework which has been discussed and analyzed extensively in the literature, see {\it e.g.}~\cite{Akal:2020twv, Akal:2022qei}. In this part, we will discuss the essential ingredients which are relevant for the latter part where we discuss our results. Consider a two-dimensional CFT in flat Minkowski geometry:
\begin{eqnarray}
ds^2 = - dt^2 + dx^2 = - du dv \ , \quad u = t - x \ , \quad v = t + x \ . \label{muv}
\end{eqnarray}
The ground-state of this CFT has a vanishing energy-momentum tensor expectation value. A mirror is a physical boundary from which the CFT degrees of freedom will ``bounce-off", preserving conformal invariance. This is achieved by imposing conformal boundary conditions on the CFT stress-tensor, in terms of a boundary CFT (BCFT) framework. Such boundary conditions can equivalently be thought of as boundary states in the CFT.\footnote{Note that, it is generally difficult to explicitly construct boundary states in a CFT, even though in examples such as free fields, Ising CFT, explicit results are known (see {\it e.g.}~\cite{Brehm:2015lja, Brehm:2015plf}).} The simplest mirror is described by $x=0$ and therefore the corresponding physical space is determined by $x>0$. In the ground state of this BCFT, the conformal boundary condition is trivially satisfied, since $\langle T_{uu}\rangle =\langle T_{vv}\rangle =0$.
%in which only right-moving CFT quanta exist that produces a non-vanishing energy flux $T_{uu}$ in the vacuum. We will remark on the possible physical interpretation of this non-vanishing flux momentarily. 

On the other hand, a time-dependent boundary, by definition, would be described by a profile of the type: $x = Z(t)$. In light cone coordinates, this can be expressed as $F(u,v)=0$ or equivalently as $v=p(u)$. Since the motivation behind the moving mirror profile is to mimic a black hole background, in which the ground state contains radiation from the black hole, instead of imposing conformal boundary condition, one imposes the so-called {\it radiative} {\it boundary} {\it conditions}, in which the out-going radiation $\langle T_{uu}\rangle \neq 0$, while $\langle T_{vv}\rangle$ vanishes. A simple way to achieve this boundary condition is to first start with a BCFT in its ground state described by coordinates $\{\tilde{t}, \tilde{x}\}$, where the usual conformal boundary condition is imposed on the static mirror at $\tilde{x}=0$,\footnote{Correspondingly, the physical space is described by $\tilde{x}>0$.} which corresponds to $\tilde{u} = \tilde{v}$, where:  
%and thereafter use the usual BCFT techniques to analyze the CFT dynamics in the presence of a static boundary condition at $\tilde{x}=0$.
%conformal boundary    we can first conformally map to a $\{\tilde{t}, \tilde{x}\}$ coordinate system where the mirror is described by $\tilde{x}=0$\footnote{Correspondingly, the physical space is described by $\tilde{x}>0$.} and thereafter use the same BCFT technique to analyze the CFT dynamics in the presence of a static boundary condition at $\tilde{x}=0$. Eventually, we revert back the results obtained in the tilde-coordinates in terms of the $\{u,v\}$ coordinates (as defined in (\ref{muv})). More explicitly, given the $\{t, x\}$ coordinate system of (\ref{muv}) (correspondingly the $\{u, v\}$ coordinates), we define a new coordinates:
%
\begin{eqnarray}
\left( \tilde{u}, \tilde{v} \right) = \left( \tilde{t} - \tilde{x}, \tilde{t} + \tilde{x} \right) \ , 
\end{eqnarray}
%
%in which the mirror is located at $\tilde{x} = 0 $ which corresponds to $\tilde{u} = \tilde{v}$. To excite the right-moving CFT degrees of freedom, we consider the following chiral conformal transformation:
%
To obtain the {\it radiative} {\it boundary} {\it condition}, one then relates $u$ and $v$ with $\tilde{u}$ and $\tilde{v}$ via a chiral conformal transformation
\begin{eqnarray}
\tilde{u} = p(u ) \ , \quad \tilde{v} = v \quad \implies \quad t + Z(t) = p \left( t - Z(t) \right) \ ,  \label{cconf}
\end{eqnarray}
where, as alluded above, $x = Z(t)$ is the profile of the mirror. As remarked in \cite{Akal:2022qei}, the above equation is equivalent to the following:
\begin{eqnarray}
\frac{p(u ) - u}{2} = Z \left( \frac{u + p(u)}{2}\right) \ .
\end{eqnarray}

As mentioned above, eqn. (\ref{cconf}) is a chiral conformal transformation.\footnote{In terms of the transformations defined on the complex-plane, using the standard $\{z, {\bar z}\}$ coordinates, chirality is realized by {\it e.g.}~transforming $z$ while keeping ${\tilde z}$  same.} Under this transformation the energy-momentum changes by a Schwarzian derivative contribution to:
\begin{eqnarray}
T_{uu} = \frac{c}{24 \pi} \left( \frac{3}{2} \left(\frac{p''(u)}{p'(u)} \right)^2 -  \frac{p'''(u)}{p'(u)} \right) \ , \quad T_{vv} = 0 \ , \quad T_{uv} = 0 \ , \label{emu}
\end{eqnarray}
which yields a non-vanishing contribution for a generic choice of the function $p(u)$, as desired. Also,  
%Note that, there are two possible interpretations of this observation: First, the energy-flux is explicitly defined in terms of the $u$ coordinate, which is a null-direction in the original $\{t, x\}$-plane. Thus, one interpretation of $T_{uu}$ as a physical energy-flux is in terms of a quantization in the CFT where $u$ is the time-direction. {\bf is there any qualifier to this statement?} Alternatively, we also have:
%
\begin{eqnarray}
T_{tt} = T_{uu} + T_{vv} - 2 T_{uv} = T_{uu} \ , 
\end{eqnarray}
using eqn. (\ref{emu}). Thus, a generic function $p(u)$ yields a non-vanishing $T_{tt}$. 
%{\bf what we eventually see in OTOC, or in EE growth, can also be interpreted in two ways?} 

Note that, when $p(u) \in SL(2,R)$, $T_{uu} = 0$. The corresponding conformal transformations are of three types:
\begin{eqnarray}
p(u) = u + \alpha \ , \quad p(u) = \lambda u \ , \quad p(u) = - \frac{\kappa}{u} \ ,
\end{eqnarray}
where $\{\alpha, \lambda, \kappa\}$ are real constants and the three choices above correspond to translation, dilatation and inversion. Using eqn. (\ref{cconf}), these cases correspond to:
\begin{eqnarray}
Z(t) = \frac{\alpha}{2} \ , \quad Z(t) = \frac{\lambda - 1}{\lambda+1} t \ , \quad Z(t) = \pm \sqrt{t^2 + \kappa} \ . \label{mirrorZ}
\end{eqnarray}
These correspond to mirrors with a uniform position ({\it i.e.}~completely static), a uniform velocity, and a uniform acceleration- the Rindler profile, respectively. 
%{\bf how do we easily see the Rindler case?} 
Furthermore, in the limit $t\gg \kappa$, $Z(t)$ approaches uniform velocity, and in the limit $t \ll \kappa$, $Z(t)$ approaches a constant mirror profile. Therefore, in order to observe a non-trivial energy flux, one must consider a non-uniform accelerating mirror.\footnote{Note that, it is now easy to construct mirror trajectories that correspond to a non-uniform acceleration. For example, a profile of the following type:
\begin{eqnarray}
Z(t) = \left( \sum_{i=0}^N t^i \right)^a \ , 
\end{eqnarray}
where $a$ and $N$ are real constants, takes us out of the $SL(2,R)$-class. 
} %{\bf Should we provide a picture?}

A few comments are in order. Using the mirror position described by $\tilde{u} = \tilde{v}$, on the mirror $2 t = u + \tilde{u} = u + p(u) $ and $2 x = \tilde{u} - u = p(u) - u $. Thus, the induced metric on the mirror is given by \cite{Akal:2022qei}
\begin{eqnarray}
ds_{\rm mirror}^2 = - p'(u) \left( \frac{2}{1+ p'(u) }\right)^2 dt^2  \ .
\end{eqnarray}
Thus, the mirror is spacelike if $p'(u) < 0$, timelike if $p'(u)>0$ and null if $p'(u) =0$, or $p'(u)\rightarrow \infty$. It is now easy to check that for the constant static profile, the mirror is always timelike. For the uniform velocity profile, it is timelike, spacelike, or null if $\lambda>0$, $\lambda<0$ or $\lambda = 0$. For the constant acceleration profile, when $\kappa>0$ the mirror is timelike but asymptotes to a null curve in the limit $u \to \infty$. These statements are also nicely visible in the corresponding profile functions $Z(t)$ given in eqn (\ref{mirrorZ}). On physical grounds, we should demand $p'(u)>0$.

%Typically one places the mirror at $x=0$, so the physical space is $x>0$, then let's assume the mirror follows the trajectory $x=Z(t)$, physical space becomes $x-Z(t)>0$. The simplest way to analyse moving mirror is by mapping the time dependent mirror trajectory to static case by conformal transformation.

%Let's use the null coordinates $v=t+x$ and $u=t-x$ to describe the mirror trajectory. 
%We consider a conformal transformation $\tilde{u}=p(u)$ and $\tilde v=v$, where $\tilde v=\tilde t+\tilde x$ and $\tilde u=\tilde t-\tilde x$, such that the mirror trajectory becomes static $\tilde{x}=0$. We can determine the conformal transformation $p(u)$ from 
%\begin{equation}
%t+Z(t)=p(t-Z(t))
%\end{equation}

%%%%%%%%%%%%%%%%%%%%%%%%%%%
\subsection{Escaping mirror}\label{subsec:escaping}
%%%%%%%%%%%%%%%%%%%%%%%%%%%

\begin{figure}[ht]
\centering
\begin{tikzpicture}[scale=0.7]
\draw [ ->] (3,-3) -- (3,3) node at (3,3.3) {$ t$};
\draw [ ->] (0,0) -- (6,0) node[right] {$x$};
\draw [dashed, ->] (0.5,-2.5) -- (5.5,2.5) node at (5.7,2.7) {$v$};
\draw [dashed, ->] (5.5,-2.5) -- (0.5,2.5) node at (0.4,2.7) {$u$};
%\draw[scale=0.5,domain=-3:3,smooth,variable=\x]   plot (\x,{tan(\x r)});
\draw[red,ultra thick] (2.85,-3) arc (0:48:7.3);
\end{tikzpicture}
\caption{Escaping mirror profile, described by the function $p(u)=-\beta \log\left(1+e^{-u/\beta}\right)$.}
\label{fig:mesh1}
\end{figure}
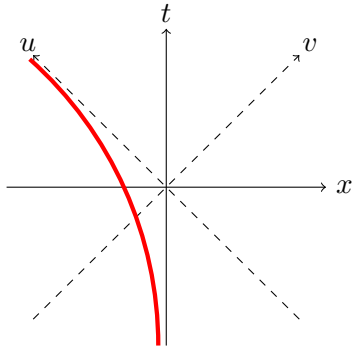

Let us begin with the ``escaping mirror" profile \cite{Akal:2020twv,Akal:2021foz}. The profile is given by
\begin{eqnarray}\label{eq:escaping}
v=p(u)=-\beta \log\left(1+e^{-u/\beta}\right) \ ,
\end{eqnarray}
where $\beta$ is a real positive parameter. Note that, the following limits are instructive:
\begin{equation}\label{eq:escap_inf}
\lim_{u \to \infty} p(u) = -\beta e^{-u/\beta} \,
\end{equation}
\begin{equation}
\lim_{u \to - \infty} p(u) = u \,
\end{equation}
where the latter ($u \to - \infty$ limit) corresponds to a static mirror. Thus, the early time behaviour corresponds to an $SL(2,R)$-class mirror profile and will contribute a vanishing energy flux; while the $u \to \infty$ limit will contribute a non-vanishing value. The corresponding mirror profile is shown in fig. \ref{fig:mesh1}.
% - holographically, it models constant Hawking radiation from an eternal black hole.
%\begin{center}
%\includegraphics[scale=0.6]{image}
%\end{center}

%Escaping mirror profile is given by $v=p(u)=-\beta \log\left(1+e^{-u/\beta}\right)$.

Note, furthermore, that $p'(u) = (1+ e^{u/\beta})^{-1}$. This is a strictly positive function and therefore the mirror is timelike. Also, note that $\lim_{u\to -\infty} p'(u) \to 1$ and $\lim_{u\to \infty} p'(u) \to e^{-u/\beta}$. Thus, at past infinity, the mirror is timelike while at the future infinity, the mirror asymptotes to a null curve. Both these features are explicitly visible in fig. \ref{fig:mesh1}. 

\begin{figure}[!h]
\centering
\begin{tikzpicture}[scale=0.7]
\fill[fill=yellow!20,] (0.5,-3)-- (2.85,-3) arc (0:48:7.3) -- (0.5,-3);
\fill[fill=green!40] (6,-3)--(2.85,-3) arc (0:48:7.3)--(2,4)--(6,4);
\draw [ ->] (3,-3) -- (3,4) node at (3,4.3) {$ t$};
\draw [ ->] (0.5,0) -- (6,0) node[right] {$x$};
\draw [dashed, ->] (0.5,-2.5) -- (6,3) node at (6.2,3) {$v$};
\draw [dashed, ->] (5.5,-2.5) -- (0.5,2.5) node at (0.4,2.7) {$u$};
\draw [->,very thick,green] (2.4,-0.57)--(6,3);
\draw[red,ultra thick] (2.85,-3) arc (0:48:7.3);
\draw [->,blue,very thick] (1.6,1.1)--(4.5,4);
\draw [->,magenta,very thick] (.5,2.5)--(2,4);
%%%
%%%%
\draw[<->, very thick] (7.4,0)--(8.2,0);
\fill[fill=yellow!20,] (12,0)-- (9,-3) -- (12,-3);
\fill[fill=green!40] (12,0)--(15,3) --(15,-3)--(12,-3);
\draw [ ->] (12,-3) -- (12,3) node at (12,3.3) {$\tilde{t}$};
\draw [ ->] (9,0) -- (15,0) node[right] {$\tilde{x}$};
\draw [dashed, ->] (9,-3) -- (15,3) node at (15.2,3.2) {$\tilde{v}$};
\draw [dashed, ->] (15,-3) -- (9,3) node at (8.8,3.2) {$\tilde{u}$};
\draw[red,ultra thick] (12,0) -- (12,-3);
\draw [<->,thick,blue] (3.1,-0.1)--(2.5,-0.7) node at (3.3,-0.5) {\tiny{$\beta \log 2$}};
\draw [->, very thick, magenta] (12,0) -- (15,3);
\draw [->, very thick, blue] (12,-.75) -- (15,2.25);
\draw [->, very thick, green] (12,-1.5) -- (15,1.5);
\filldraw  (1.1,1.7) circle (3pt) node[right] {$P_1$} ;
\filldraw  (1.97,0.5) circle (3pt) node[right] {$P_2$} ;
\filldraw  (4.5,-0.5) circle (3pt) node[right] {$P(u,v)$} ;
\filldraw  (12,-0.37) circle (3pt) node[left] {$P_1$} ;
\filldraw  (12,-1.12) circle (3pt) node[right] {$P_2$} ;
\filldraw  (13.5,-1) circle (3pt) node[right] {$P(\tilde{u},\tilde{v})$} ;
\filldraw  (10.5,-1) circle (3pt) node[left] {$P_r(\tilde{v},\tilde{u})$} ;
\end{tikzpicture}
\caption{Static mirror from escaping mirror using conformal mapping. In the figure, we have explicitly depicted three operator insertions: two identical operators, denoted by $V$, inserted on the boundary of the mirror at points $P_1$ and $P_2$, and another operator, denoted by $W$, inserted in the bulk of the physical space at point $P$. The bulk operator $W$ creates an image in the resulting BCFT and therefore a three-point function of $V$ and $W$ becomes a four-point function in the CFT.}
\label{fig:escaping}
\end{figure}
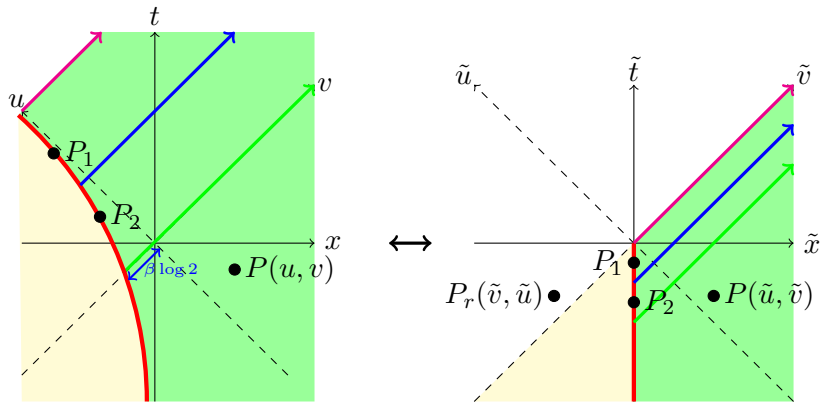

%\subsubsection*{Mapping the escaping mirror to a static mirror}
Let us now consider the conformal transformation that maps the moving mirror to a static mirror. As we reviewed above, under $\tilde{u}=p(u)$ and $\tilde{v}=v$, the escaping mirror $x=Z(t)$ gets mapped to a static mirror $\tilde{x}=0$. This is pictorially demonstrated in fig. \ref{fig:escaping}. Additionally, in these figures, we have explicitly inserted three primary operators. Specifically, we placed the same operator $V$ at two points $P_1$ and $P_2$ on the mirror boundary, and another operator $W$ in the bulk at the point $P$. We will discuss the correlation function of these operators and will connect the result to an out-of-time-order correlator (OTOC) momentarily. 

Before delving into the OTOCs, let us review the basic features of the moving mirror. The corresponding energy-momentum can be calculated using (\ref{emu}) and the specific function of $p(u)$. This yields \cite{Akal:2021foz}:
\begin{eqnarray}
T_{uu} = \frac{c}{48 \pi \beta^2} \left( 1 - \frac{1}{\left( 1+ e^{u/\beta}\right)^2}\right)  \ , 
\end{eqnarray}
which has the following instructive limits: $\lim_{u\to- \infty}T_{uu} \to 0$ and $\lim_{u\to +\infty}T_{uu} \to c/ (48 \pi \beta^2)$. The latter is precisely the energy-momentum expectation value in a thermal state of inverse temperature $(2\pi \beta)$.\footnote{Note that, the large and the small $u$ limits are precisely defined in terms of the parameter $u/\beta$. For example, it is easy to observe that in the limit $u \to \infty$ and $\beta \to \infty$, such that $u/\beta$ is fixed, the energy flux $T_{uu}$ never settles to the thermal value. } Physically, this thermal energy flux is identified with the constant Hawking radiation from an eternal black hole in two dimensions. As further evidenced in \cite{Akal:2021foz}, by considering a massless free CFT in the moving mirror background, one obtains a finer matching of the system with radiation from an eternal black hole in terms of Bogoliubov coefficients and thermal spectrum.

We will now discuss the calculation of a certain class of OTOCs, which are referred to as $b$-OTOCs in \cite{Das:2019tga}. While the free CFT thermal result is very encouraging, it is a purely kinematic answer based on a two-point function in a CFT and therefore does not capture dynamical information of the system. By analyzing the OTOCs, we would like to probe finer details of the dynamics of the corresponding CFT. 

Correlation function in a Lorentzian quantum field theory can be obtained from its Euclidean counterpart through analytic continuation. To achieve this operationally, one begins with a Euclidean correlator with operators inserted at small and purely imaginary times $i\epsilon_i$. Then analytically continue the time coordinate to the desired Lorentzian value $t_i+i\epsilon_i$ and finally take $\epsilon_i\rightarrow 0$. Different operators in the Lorentzian correlator will be ordered according to the order in which $\epsilon_i$'s are taken to zero, {\it e.g.} $\langle O(t_1)O(t_2)\dots O(t_n)\rangle$ will correspond to the ordering $\epsilon_1<\epsilon_2<\dots<\epsilon_n$ \cite{Osterwalder:1973dx,Haag:1992hx}.

In CFTs, the two and three-point functions exhibit a simple structure determined by conformal symmetry, up to an overall constant. However, in the context of BCFTs, the reduced symmetry leads to nontrivial and interesting features in these correlation functions. Specifically, the conformal Ward identities imply that the $n$-point function in a BCFT behaves like a $2n$-point function of a holomorphic bulk CFT under conformal transformations \cite{Cardy:1984bb}. While the two-point function in a BCFT resembles a four-point function in a holomorphic bulk CFT, it can not be an OTOC and therefore not of much use to us. With this motivation, we will consider the analytic structure of a three-point OTOC with two operators on the boundary and one in the bulk.

%\subsubsection*{Conformal cross ratio}

We begin with operator insertions shown in fig. \ref{fig:escaping}. In terms of $(\tilde{u},\tilde{v})$, the coordinates of the boundary points satisfy $\tilde{u}=\tilde{v}$. Let us assign the boundary points $P_1$ and $P_2$ the following coordinates : $P_1\equiv (\tilde{v}_1,\tilde{v}_1)$ and $P_2\equiv (\tilde{v}_2,\tilde{v}_2)$. %\footnote{There is another choice for the points $P_1$ and $P_2$, $P_1\equiv (\tilde{u}_1,\tilde{u}_1)$ and $P_2\equiv (\tilde{u}_2,\tilde{u}_2)$. We can compute cross ratio $\eta$ with this choice as well, however, the results will not change.}
Let us also assign the point in the bulk $P$ the following coordinates $P\equiv(\tilde{u},\tilde{v})$. Using Cardy's {\it doubling trick} \cite{Cardy:1984bb}, the above three-point function will be the same as that of a four-point function with one extra operator $W$ at the reflected point $P_r$ with coordinates $P_r\equiv(\tilde{v},\tilde{u})$.
The conformal cross ratio is
\begin{equation}\label{eq:cross_ratio}
\begin{split}
\eta&=\frac{(\tilde{u}-\tilde{v})(\tilde{v}_1-\tilde{v}_2)}{(\tilde{u}-\tilde{v}_1)(\tilde{v}-\tilde{v}_2)} \ . 
\end{split}
\end{equation}
As discussed above, we will now analytically continue to  $\tilde{t}\rightarrow \tilde{t}+i\,\tilde{\epsilon}$.  
%{\bf in terms of $t$ or $\tilde{t}$? Also, $t$ is already Lorentzian here.} {\color{red}\bf In terms of $\tilde{t}$, the operators in OTOCs are getting ordered according to the analytic continuation in $(\tilde{x},\tilde{t})$ coordinates not in $(x,t)$ coordinates.} 
%Then keeping $\tilde{\epsilon}$ fixed, we will take $\tilde{t}$ to the desired Lorentzian value. Different operators in the OTOC will get ordered according to the value of $\tilde{\epsilon}$.
We will be doing this analytic continuation in $(\tilde{u},\tilde{v})$ coordinates (see appendix \ref{app:analytic} for discussion of analytic continuation in $(u,v)$ coordinates). Therefore, in $(\tilde{u},\tilde{v})$ coordinates, we have:
\begin{equation}
\begin{split}
P_1&\equiv (\tilde{v}_1+i\,\tilde{\epsilon}_1, \tilde{v}_1+i\,\tilde{\epsilon}_1) \ ,\qquad P_2\equiv (\tilde{v}_2+i\,\tilde{\epsilon}_2, \tilde{v}_2+i\,\tilde{\epsilon}_2) \ ,\\
P&\equiv (\tilde{u}+i\,\tilde{\epsilon}_0, \tilde{v}+i\,\tilde{\epsilon}_0) \ , \quad~\quad~ P_r\equiv (\tilde{v}+i\,\tilde{\epsilon}_0, \tilde{u}+i\,\tilde{\epsilon}_0) \ .
\end{split}
\end{equation}
The conformal cross-ratio is given by
\begin{equation}
\begin{split}
\eta&=\frac{(\tilde{u}+i\,\tilde{\epsilon}_0-\tilde{v}-i\,\tilde{\epsilon}_0)(\tilde{v}_1+i\,\tilde{\epsilon}_1-\tilde{v}_2-i\,\tilde{\epsilon}_2)}{(\tilde{u}+i\,\tilde{\epsilon}_0-\tilde{v}_1-i\,\tilde{\epsilon}_1)(\tilde{v}+i\,\tilde{\epsilon}_0-\tilde{v}_2-i\,\tilde{\epsilon}_2)}\ .
\end{split}
\end{equation}
We will take the two boundary points $P_1$ and $P_2$ at a same point $P_1\equiv P_2$ which implies $\tilde{v}_1=\tilde{v}_2$. The conformal cross ratio becomes:
\begin{equation}\label{eq:etaescaping}
\begin{split}
\eta&=\frac{(\tilde{u}-\tilde{v})i\,\tilde{\epsilon}_{12}}{(\tilde{u}-\tilde{v}_1+i\,\tilde{\epsilon}_{01})(\tilde{v}-\tilde{v}_1+i\,\tilde{\epsilon}_{02})}\\
&=\frac{\big(p(u)-v\big)i\,\tilde{\epsilon}_{12}}{\big(p(u)-v_1+i\,\tilde{\epsilon}_{01}\big)\big(v-v_1+i\,\tilde{\epsilon}_{02})}\\
&=i\,\tilde{\epsilon}_{12}\,\frac{-\beta\log\left(1+e^{-\frac{u}{\beta}}\right)-v}{\left[-\beta\log\left(1+e^{-\frac{u}{\beta}}\right)-v_1+i\,\tilde{\epsilon}_{01}\right]\left[v-v_1+i\,\tilde{\epsilon}_{02}\right]} \ .
\end{split}
\end{equation}
We will set $v=0$ for the bulk operator $W$. In terms of $(t,x)$ it implies $x=-t$, therefore $u=2t$ for the bulk operator. Setting $v=0$, we will study behaviour of b-OTOC as we vary $u$ from $0$ to $\infty$.
\subsection*{Case-1: $v=0, u\rightarrow\infty$}
\begin{equation}
\eta=i\tilde{\epsilon}_{12}\frac{-\beta\, e^{-u/\beta}}{v_1^2} \ .
\end{equation}

\subsection*{Case-2: $v=0, u=0$}
\begin{equation}
\eta=i\tilde{\epsilon}_{12}\frac{\beta \log 2}{(\beta\log 2+v_1)(-v_1)} \ .
\end{equation}
$-v_1>0$, and let's assume $\beta\log 2+v_1>0$. The above assumptions imply that the points $P_1$ and $P_2$ have to lie in the quadrant $(u>0,v<0)$ in the $(u,v)$ plane. So we see that $\eta$ approaches $0$ from the opposite directions for Case-1 and Case-2.

\subsection*{Case-3: $v=0, u=u_1=p^{-1}(v_1)$}
\begin{equation}
\begin{split}
\eta&=i\,\tilde{\epsilon}_{12}\frac{-\beta\log\left(1+e^{-\frac{u_1}{\beta}}\right)}{\left[-\beta\log\left(1+e^{-\frac{u_1}{\beta}}\right)-v_1+i\,\tilde{\epsilon}_{01}\right][-v_1]}\\
&=-\frac{\tilde{\epsilon}_{12}}{\tilde{\epsilon}_{01}}=1+\frac{\tilde{\epsilon}_{20}}{\tilde{\epsilon}_{01}} \ .
\end{split}
\end{equation}

The cross ratio $\eta$ goes to zero from the opposite direction in the limit $u\rightarrow 0$ and $u\rightarrow\infty$. 
For the intermediate point $u=u_1$, in the out-of-time-ordered case, $\tilde{\epsilon}_2>\tilde{\epsilon}_0>\tilde{\epsilon}_1$ or, $\tilde{\epsilon}_1>\tilde{\epsilon}_0>\tilde{\epsilon}_2$, it crosses the real axis after $\eta=1$ branch point. But for the time-ordered case, it crosses the real axis before the $\eta=1$ branch point. This behaviour is similar to what was obtained in \cite{Das:2019tga}, with one interesting difference, which is that the ordering here, refers to the ordering of operators in the static frame. As we discuss in detail in appendix \ref{app:analytic}, in the moving mirror frame {\it i.e.} in the ($u,v$) coordinates, the ordering becomes $\frac{\epsilon_2}{p'(u_1)}>\epsilon_0>\frac{\epsilon_{1}}{p'(u_1)}$. Here $p'(u)$ is the Weyl scaling factor of the metric at the location of mirror $(d\tilde{u}d\tilde{v}=\frac{1}{p'(u)}dudv)$ and $u_1$, as mentioned previously, is the location of the boundary operators on the mirror. So, the operator ordering, for boundary operators, in the mirror frame, is effectively decided by $\frac{\epsilon_i}{p'(u)}$. It would be nice to better understand the physical reason behind this scaling of the boundary $\epsilon$'s. One should note that such scale factors also appear in the presence of a moving interface profile, where the   
transmission and reflection coefficients, as computed in the moving frame, are off from their static frame counterparts by such scaling factors (see {\it e.g.} eqn. (3.23) of \cite{Biswas:2024xut}).   
%Instead of doing analytic continuation in $(\tilde{u},\tilde{v})$ coordinates, we can also do the analytic continuation in $(u,v)$ coordinates.

%Instead of doing analytic continuation in $(\tilde{u},\tilde{v})$ coordinates, we can also do the analytic continuation in $(u,v)$ coordinates. 

So far, the discussion is purely kinematical, {\it i.e.}~holds for any two-dimensional CFT. To determine the explicit form of the four-point function, we will now need to invoke dynamical input of the specific theory. Let us begin with large-$c$ CFTs, assuming the identity block dominance in the corresponding four-point function. The explicit form of the identity block is given, as a function of the cross-ration, by
\begin{equation}
\mathcal{F}_0(\eta)\approx \left(\frac{\eta}{1-(1-\eta)^{1-\frac{12 h_w}{c}}}\right)^{2h_v} \ .
\end{equation}
The above formula is valid at large central charge $c$, with $\frac{h_w}{c}$ fixed and small, and $h_v$ fixed and large. We take our bulk operator to be $W$ with scaling dimension $h_w$ and the boundary operator to be $V$ with scaling dimension $h_{v}$.

In the OTO case, $\eta$ crosses the branch point at $\eta=1$ and therefore, in the second Riemann sheet $ 1- \eta \to (1 - \eta) e^{-2\pi i}$. Taking $\eta$ small, we obtain:
\begin{equation}\label{eq:vac_vir}
\begin{split}
\mathcal{F}_0(\eta)%&\approx \left(\frac{\eta}{1-(1-\eta)^{1-\frac{12 h_w}{c}}}\right)^{2h_v}\\
&=\left(\frac{\eta}{1-(1-\eta)^{1-\frac{12 h_w}{c}}e^{-2\pi i\left(1-\frac{12 h_w}{c}\right)}}\right)^{2h_v}\\
&=\left(\frac{\eta}{1-\left[1-\left(1-\frac{12 h_w}{c}\right)\eta\right]e^{\frac{24\pi i h_w}{c}}}\right)^{2h_v}\\
&=\left(\frac{\eta}{1-\left[1-\eta\right]\left[1+\frac{24\pi i h_w}{c}\right]}\right)^{2h_v}\\
&=\left(\frac{1}{1-\frac{24\pi i h_w}{c\eta}}\right)^{2h_v}\\
&=\left(\frac{1}{1+\frac{24\pi  h_w v_1^2}{c\, \tilde{\epsilon}_{12}\beta}e^\frac{u}{\beta}}\right)^{2h_v} \approx  1 - \frac{48\pi  h_v h_w v_1^2}{c \tilde{\epsilon}_{12} \beta} e^{u/\beta} + \cdots \,.
\end{split}
\end{equation}
We see exponential behaviour with Lyapunov index $\lambda=\frac{1}{\beta} = 2\pi T$. This is the maximal Lyapunov exponent and is therefore maximally chaotic as described by OTOCs. 

%{\bf Add comments/discussions on how free CFT or Ising CFT does not display any chaotic feature at all.}

On the other hand, Ising CFT and Orbifold CFT do not display any chaotic features at all. For Ising CFT (see appendix B of \cite{Roberts:2014ifa})
\begin{equation}\label{eq:isingeta}
f_{\sigma\epsilon}(\eta,\bar{\eta})= \left(\frac{2-\eta}{2\sqrt{1-\eta}}\right)\left(\frac{2-\bar{\eta}}{2\sqrt{1-\bar{\eta}}}\right)\,.
\end{equation}
As we take $u\rightarrow\infty$, both $\eta,\bar{\eta}\rightarrow 0$. $\eta$ crosses the real axis after the brunch point at $\eta=1$. From \eqref{eq:isingeta}, we get
\begin{equation}
\begin{split}
f_{\sigma\epsilon}(\eta,\bar{\eta})&=\frac{2-\eta}{2(1-\frac{1}{2}\eta)e^{-i\pi}}\,\frac{2-\bar\eta}{2(1-\frac{1}{2}\bar\eta)}\\
&=-1\,.
\end{split}
\end{equation}
%Ising CFT does not show chaotic behaviour with escaping mirror boundary condition.\\\\
%
For the orbifold CFT $(T^2)^2/Z_2$ (see eqn. (3.29) of \cite{Caputa:2017rkm}),
\begin{equation}
\begin{split}
F_2(\eta,\bar{\eta})&\simeq \frac{i\pi}{2\log\left(\frac{\eta}{16}\right)}\,.\\
%&=\frac{i\pi}{2\log\left(-i\frac{\tilde{\epsilon}_{12}\beta}{16 v_1^2}e^{-\frac{u}{\beta}}\right)}
\end{split}
\end{equation}
Therefore, in the $u\rightarrow \infty$ limit, we will get
\begin{equation}
\begin{split}
F_2(\eta,\bar{\eta})
&=\frac{i\pi}{2\log\left(-i\frac{\tilde{\epsilon}_{12}\beta}{16 v_1^2}e^{-\frac{u}{\beta}}\right)}\approx -\frac{i\pi\beta}{2u}\,.
\end{split}
\end{equation}
Thus for the orbifold CFTs, the dependence on $u$ is polynomial.

%%%%%%%%%%%%%%%%%%%%%%
\subsection{Kink mirror}\label{subsec:kink}
%%%%%%%%%%%%%%%%%%%%%%

Let us now consider the kink mirror \cite{Akal:2020twv,Akal:2021foz} which models an evaporating black hole formed from collapse. The corresponding mirror profile is given by
\begin{equation}\label{eq:kinkmirror}
v=p(u)=-\beta\log\left(1+e^{-\frac{u}{\beta}}\right)+\beta \log\left(1+e^{\frac{u-u_0}{\beta}}\right) \ , 
\end{equation}
where, we now have two real positive parameters $u_0$ and $\beta$. As before, let us inspect some instructive limits:
\begin{eqnarray}
&& \lim_{u \to \pm \infty} p(u) = u \ , \\
&& \lim_{u \to u_0 } p(u)  = - \beta \log \left( 1 + e^{- u_0/\beta}\right) + \beta \log 2 \ , 
\end{eqnarray}
%
%{\color{red}\bf The second term in the above equation should be $\beta\log 2$ instead of $\beta e^{\frac{u - u_0}{\beta}}$.}\\
which implies that in the $u \to \pm \infty$ limit, the mirror profile belongs to an $SL(2,R)$-class and therefore the corresponding energy flux vanishes.\footnote{In fact, in both these limits, the mirror becomes a static one.} In the $u \to u_0$ limit, however, non-trivial flux is created. Given the profile, we also obtain:
\begin{eqnarray}
p'(u) = \left ( 1 + e^{u/\beta}\right)^{-1} + \left ( 1 + e^{(u - u_0)/\beta}\right)^{-1} > 0 \ , 
\end{eqnarray}
and therefore the mirror is timelike. Unlike the escaping mirror, there is no limit where $p'(u)$ can asymptotically vanish. It is now straightforward to obtain the corresponding expectation value of $T_{uu}$. Furthermore, the kink mirror models the formation and evaporation of a black hole, which can be observed by comparing the Bogoliubov coefficients of a free scalar theory in a two-dimensional collapsing and evaporating black hole with the same in a kink mirror geometry \cite{Akal:2020twv}. Note, furthermore, that in \cite{Akal:2020twv}, a fine-grained entropic computation has been shown to exhibit the Page curve associated to the unitary evolution of a quantum black hole.

\begin{figure}[!h]
\centering
\begin{tikzpicture}[scale=0.6]
\fill[fill=green!40] (4,4)--plot [smooth] coordinates {(-1.9316, 3.9683) (-1.909, 3.690) (-1.8711, 3.3288) (-1.8185, 2.9814) (-1.7495, 2.6504) (-1.662, 2.337) (-1.5569, 2.0430) (-1.4344, 1.7655) (-1.2978, 1.5021) (-1.1514, 1.2485) (-1, 1) (-0.8485, 0.7514) (-0.7021, 0.4978) (-0.5655, 0.2344) (-0.4430, -0.0430) (-0.3374, -0.3374) (-0.2504, -0.6504) (-0.1814, -0.9814) (-0.1288, -1.3288) (-0.0901, -1.6901) (-0.0622, -2.0622) (-0.0425, -2.4425) (-0.0289, -2.8289) (-0.0196, -3.2196) (-0.0132, -3.6132) (-0.0098, -3.9098)}--(4,-4);
\fill[fill=yellow!20,] (-4,4)--plot [smooth] coordinates {(-1.9316, 3.9683) (-1.909, 3.690) (-1.8711, 3.3288) (-1.8185, 2.9814) (-1.7495, 2.6504) (-1.662, 2.337) (-1.5569, 2.0430) (-1.4344, 1.7655) (-1.2978, 1.5021) (-1.1514, 1.2485) (-1, 1) (-0.8485, 0.7514) (-0.7021, 0.4978) (-0.5655, 0.2344) (-0.4430, -0.0430) (-0.3374, -0.3374) (-0.2504, -0.6504) (-0.1814, -0.9814) (-0.1288, -1.3288) (-0.0901, -1.6901) (-0.0622, -2.0622) (-0.0425, -2.4425) (-0.0289, -2.8289) (-0.0196, -3.2196) (-0.0132, -3.6132) (-0.0098, -3.9098)}--(-4,-4);
\draw [ ->] (0,-4) -- (0,4) node at (0,4.3) {$ t$};
\draw [ ->] (-4,0) -- (4,0) node[right] {$x$};
\draw [dashed, ->] (-4,-4) -- (4,4) node at (4.2,4.2) {$v$};
\draw [dashed, ->] (4,-4) -- (-4,4) node at (-4.2,4.2) {$u$};
\draw[red,ultra thick] plot [smooth] coordinates {(-1.9316, 3.9683) (-1.909, 3.690) (-1.8711, 3.3288) (-1.8185, 2.9814) (-1.7495, 2.6504) (-1.662, 2.337) (-1.5569, 2.0430) (-1.4344, 1.7655) (-1.2978, 1.5021) (-1.1514, 1.2485) (-1, 1) (-0.8485, 0.7514) (-0.7021, 0.4978) (-0.5655, 0.2344) (-0.4430, -0.0430) (-0.3374, -0.3374) (-0.2504, -0.6504) (-0.1814, -0.9814) (-0.1288, -1.3288) (-0.0901, -1.6901) (-0.0622, -2.0622) (-0.0425, -2.4425) (-0.0289, -2.8289) (-0.0196, -3.2196) (-0.0132, -3.6132) (-0.0098, -3.9098)};
\filldraw  (-0.7,0.45) circle (3pt) node[left] {$P_1$} ;
\filldraw  (-0.45,-0.05) circle (3pt) node[left] {$P_2$} ;
\filldraw  (2,-1.2) circle (3pt) node[right] {$P(u,v)$} ;
\draw [->,blue,very thick] (-0.35,-0.35)--(4,4);
\draw [->,magenta,very thick] (-1,1)--(2,4);
\draw [<->,blue,thick] (0.1,0.1)--(-0.9,1.1) node at(-0.2,0.9) {$\frac{u_0}{2}$};
%%%%
%%%%%
\draw[<->, very thick] (5.4,0)--(6.2,0);
\fill[fill=green!40] (11,4)--(15,4)--(15,-4)--(11,-4);
\fill[fill=yellow!20,] (11,4)--(7,4)--(7,-4)--(11,-4);
\draw [ ->] (11,-4) -- (11,4) node at (11,4.4) {$\tilde{t}$};
\draw [ ->] (7,0) -- (15,0) node[right] {$\tilde{x}$};
\draw [dashed, ->] (7,-4) -- (15,4) node at (15.2,4.2) {$\tilde{v}$};
\draw [dashed, ->] (15,-4) -- (7,4) node at (6.7,4.3) {$\tilde{u}$};
\draw[red,ultra thick] (11,4) -- (11,-4);
\filldraw  (11,-0.5) circle (3pt) node[right] {$P_1$} ;
\filldraw  (11,-1) circle (3pt) node[left] {$P_2$} ;
\filldraw  (12.5,-1) circle (3pt) node[right] {$P(\tilde{u},\tilde{v})$} ;
\filldraw  (9.5,-1) circle (3pt) node[left] {$P_r(\tilde{v},\tilde{u})$} ;
\draw [->,magenta,very thick] (11,0)--(15,4);
\draw [->,blue,very thick] (11,-1.5)--(15,2.5);
\end{tikzpicture}
\caption{Static mirror from kink mirror using conformal mapping. We have also explicitly shown operator insertions in the corresponding CFT. These are primary operators inserted at two boundary points $P_{1,2}$ and at one bulk point denoted by $P$.}
\label{fig:kink}
\end{figure}
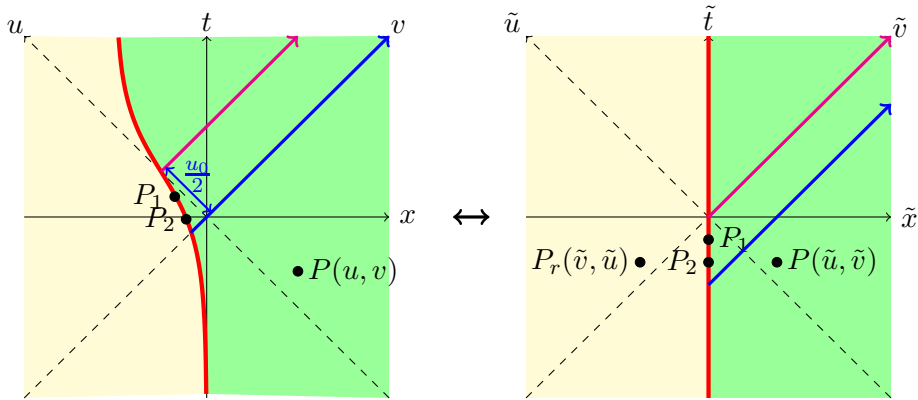

Our goal here is to analyze the corresponding $b$-OTOCs and understand the physics of scrambling. Towards this, we arrange for a similar operator insertion that we have discussed in the previous section. This is pictorially shown in fig. \ref{fig:kink}. We perform the usual conformal transformation $\tilde{u}=p(u)$ and $\tilde{v}=v$ to obtain the static mirror configuration. 

Let us take the bulk point to be $P\equiv(\tilde{u},\tilde{v})$, the reflected point would be $P_r\equiv(\tilde{v},\tilde{u})$. Boundary points are $P_1\equiv(\tilde{v}_1,\tilde{v}_1)$ and $P_2\equiv(\tilde{v}_2,\tilde{v}_2)$. We will fuse the two boundary points $P_1$ and $P_2$ to coincide, $P_1\equiv P_2$, which implies $\tilde{v}_1=\tilde{v}_2$. The cross ratio is defined as:
\begin{equation}
\begin{split}
\eta&=\frac{(\tilde{u}+i\,\tilde{\epsilon}_0-\tilde{v}-i\,\tilde{\epsilon}_0)(\tilde{v}_1+i\,\tilde{\epsilon}_1-\tilde{v}_2-i\,\tilde{\epsilon}_2)}{(\tilde{u}+i\,\tilde{\epsilon}_0-\tilde{v}_1-i\,\tilde{\epsilon}_1)(\tilde{v}+i\,\tilde{\epsilon}_0-\tilde{v}_2-i\,\tilde{\epsilon}_2)}\\
&=\frac{(\tilde{u}-\tilde{v})i\,\tilde{\epsilon}_{12}}{(\tilde{u}-\tilde{v}_1+i\,\tilde{\epsilon}_{01})(\tilde{v}-\tilde{v}_1+i\,\tilde{\epsilon}_{02})}\\
&=\frac{\big(p(u)-v\big)i\,\tilde{\epsilon}_{12}}{\big(p(u)-v_1+i\,\tilde{\epsilon}_{01}\big)\big(v-v_1+i\,\tilde{\epsilon}_{02})}\\
&=i\,\tilde{\epsilon}_{12}\frac{-\beta\log\left(1+e^{-\frac{u}{\beta}}\right)+\beta\log\left(1+e^{\frac{u-u_0}{\beta}}\right)-v}{\left[-\beta\log\left(1+e^{-\frac{u}{\beta}}\right)+\beta\log\left(1+e^{\frac{u-u_0}{\beta}}\right)-v_1+i\,\tilde{\epsilon}_{01}\right]\left[v-v_1+i\,\tilde{\epsilon}_{02}\right]} \ . 
\end{split}
\end{equation}
%%
%%%
%%%%
%\begin{equation}
%\begin{split}
%\eta&=\frac{(\tilde{u}-\tilde{v})(\tilde{v}_1-\tilde{v}_2)}{(\tilde{u}-\tilde{v}_1)(\tilde{v}-\tilde{v}_2)}\\
%&=\frac{(p(u)-v)(v_1-v_2)}{(p(u)-v_1)(v-v_2)}\\
%&=i\epsilon_{12}\frac{-\beta\log\left(1+e^{-\frac{u+i\epsilon_0}{\beta}}\right)+\beta\log\left(1+e^{\frac{u-u_0+i\epsilon_0}{\beta}}\right)-v-i\epsilon_0}{\left[-\beta\log\left(1+e^{-\frac{u+i\epsilon_0}{\beta}}\right)+\beta\log\left(1+e^{\frac{u-u_0+i\epsilon_0}{\beta}}\right)-v_1-i\epsilon_1\right]\left[v-v_2+i\epsilon_{02}\right]}
%\end{split}
%\end{equation}
%We take the two boundary points at the same point $P_1\equiv P_2$.
\subsection*{Case-1: $v=0,u=0$}
\begin{equation}
\begin{split}
\eta%&=i\epsilon_{12}\frac{-\beta\log 2+\beta\log\left(1+e^{-\frac{u_0}{\beta}}\right)}{\left[-\beta\log 2+\beta\log\left(1+e^{-\frac{u_0}{\beta}}\right)-v_1\right]\left[-v_1\right]}\\
&=i\,\tilde{\epsilon}_{12}\frac{\beta\log 2-\beta\log\left(1+e^{-\frac{u_0}{\beta}}\right)}{\left[\beta\log 2-\beta\log\left(1+e^{-\frac{u_0}{\beta}}\right)+v_1\right]\left[-v_1\right]}\,.
\end{split}
\end{equation}
We assume $v_1<0$, $\beta\log\left(\frac{2}{1+e^{-\frac{u_0}{\beta}}}\right)>|v_1|$.  If we set $u=0$ in the kink mirror profile \eqref{eq:kinkmirror} we get
\begin{equation}
v=-\beta\log\left(\frac{2}{1+e^{-\frac{u_0}{\beta}}}\right)\,.
\end{equation}
So, the above assumptions imply that the points $P_1$ and $P_2$ have to lie in the quadrant $(u>0,v<0)$ in the $(u,v)$ plane.

\subsection*{Case-2: $v=0$, $u=\frac{u_0}{2}-\alpha$, $\alpha\rightarrow 0$}
We can not take $u\rightarrow\infty$ limit in the kink mirror case. Instead, we will take $u=\frac{u_0}{2}-\alpha$ where $\alpha$ is a small positive number and will take $\alpha\rightarrow 0$ limit. We get
\begin{equation}
\begin{split}
\eta&=i\,\tilde{\epsilon}_{12}\frac{-\beta\log\left(1+e^{-\frac{1}{\beta}\Big(\frac{u_0}{2}-\alpha\Big)}\right)+\beta\log\left(1+e^{\frac{1}{\beta}\Big(-\frac{u_0}{2}-\alpha\Big)}\right)}{\left[-\beta\log\left(1+e^{-\frac{1}{\beta}\Big(\frac{u_0}{2}-\alpha\Big)}\right)+\beta\log\left(1+e^{\frac{1}{\beta}\Big(-\frac{u_0}{2}-\alpha\Big)}\right)-v_1\right]\left[-v_1\right]}\\
&=-i\,\tilde{\epsilon}_{12}\frac{2\alpha}{\Big(1+e^{\frac{u_0}{2\beta}}\Big)v_1^2}\,.
\end{split}
\end{equation}
So, we see that $\eta$ approaches $0$ from the opposite directions for case-1 and case-2.

%Instead of taking the above limit, if we take the limit $v=u-u_0,\, u\rightarrow\infty$, we will get
%\begin{equation}
%\begin{split}
%\eta&=i\epsilon_{12}\frac{-\beta e^{-\frac{u}{\beta}}+(u-u_0)+\beta e^{-\frac{u-u_0}{\beta}}-(u-u_0)}{u^2}\\
%\end{split}
%\end{equation}

\subsection*{Case-3: $v=0, u=u_1=p^{-1}(v_1)$}
\begin{equation}
\begin{split}
\eta&=i\,\tilde{\epsilon}_{12}\frac{-\beta\log\left(1+e^{-\frac{u_1}{\beta}}\right)+\beta\log\left(1+e^{\frac{u_1-u_0}{\beta}}\right)}{\left[-\beta\log\left(1+e^{-\frac{u_1}{\beta}}\right)+\beta\log\left(1+e^{\frac{u_1-u_0}{\beta}}\right)-v_1+i\,\tilde{\epsilon}_{01}\right]\left[-v_1\right]}\\
%&=-\frac{i\epsilon_{12}}{\left[-\beta\log\left(1+e^{-\frac{u_1+i\epsilon_0}{\beta}}\right)+\beta\log\left(1+e^{\frac{u_1-u_0+i\epsilon_0}{\beta}}\right)-v_1-i\epsilon_1\right]}\\
%&=-\frac{i\,\epsilon_{12}}{i\,\epsilon_0\left(\frac{1}{1+e^{\frac{u_1}{\beta}}}+\frac{1}{1+e^{-\frac{u_1-u_0}{\beta}}}\right)-i\epsilon_1}\\
&=-\frac{\tilde{\epsilon}_{12}}{\tilde{\epsilon}_{01}}=1+\frac{\tilde{\epsilon}_{20}}{\tilde{\epsilon}_{01}} \ .
\end{split}
\end{equation}
%where, we have defined $\tilde{\epsilon}_0=\epsilon_0\left(\frac{1}{1+e^{\frac{u_1}{\beta}}}+\frac{1}{1+e^{-\frac{u_1-u_0}{\beta}}}\right)$
%For, out of time order case, $\epsilon_2>\epsilon_{\tilde{0}}>\epsilon_1$ or, $\epsilon_1>\epsilon_{\tilde{0}}>\epsilon_2$, cross ratio $\eta>1$.
The cross ratio $\eta$ goes to zero from the opposite direction in the limit $u\rightarrow 0$ and $\alpha\rightarrow 0$. 
For the intermediate point $u=u_1$, in the out of time order case, $\tilde{\epsilon}_2>\tilde{\epsilon}_0>\tilde{\epsilon}_1$ or, $\tilde{\epsilon}_1>\tilde{\epsilon}_0>\tilde{\epsilon}_2$, it crosses the real axis after $\eta=1$ branch point. But for time-ordered case, it crosses the real axis before the $\eta=1$ branch point. As before, the discussion up to this point is completely general and applies to any two-dimensional CFT.

For a large-$c$ CFT, as before, we can again use the Identity block result. From \eqref{eq:vac_vir}, we can very easily see that as $\alpha\rightarrow0$:
\begin{equation}
\begin{split}
\mathcal{F}_0(\eta)
&=\left(\frac{1}{1-\frac{24\pi i h_w}{c\eta}}\right)^{2h_v}\\
&=\left(\frac{1}{1+\frac{24\pi  h_w}{c\,\tilde{\epsilon}_{12}2\alpha}\left(1+e^{\frac{u_0}{2\beta}}\right)v_1^2}\right)^{2h_v}\\
&=1-\frac{24\pi h_v h_w}{c\tilde{\epsilon}_{12}}\left(1+e^{\frac{u_0}{2\beta}}\right)v_1^2\left(\frac{1}{\alpha}\right)\,.
\end{split}
\end{equation}
In this case, we get polynomial behaviour instead of exponential behaviour and therefore does not contain a scrambling time-scale. 
%%%%%%%%%%%%%%%%%%%%%%%%%%%%%%%%%%%%%%%%%%%%%%%%%%%

%%%%%%%%%%%%%%%%%%%%%%%%%%%%%%%%%%%%%%%%%%%%%%%%%%%
\subsection{Escaping mirror from kink mirror}\label{subsec:escapingfromkink}
%%%%%%%%%%%%%%%%%%%%%%%%%%%%%%%%%%%%%%%%%%%%%%%%%%%

It is possible to obtain the escaping mirror from the kink mirror profile. Relatedly, this is possible if we can switch off the evaporation part of the black hole. This is certainly possible since the additional parameter $u_0$ controls the evaporation time-scale, or more precisely, the Page time. If we push away the Page time to a parametrically large distance, the eternal black hole dynamics remains and we expect to capture a scrambling dynamics.

Escaping mirror profile is given by $(\beta>0,u_0>0)$:
\begin{equation}
v=p(u)=-\beta \log\left(1+e^{-\frac{u}{\beta}}\right)+\beta\log\left(1+e^{\frac{u-u_0}{\beta}}\right) \ . 
\end{equation}
In the limit $u_0>>u$, the second term becomes small, so the kink mirror profile becomes a small perturbation around the escaping mirror:
\begin{equation}\label{eq:pert_kink}
v=p(u)=-\beta \log\left(1+e^{-\frac{u}{\beta}}\right)+\delta\,\beta\, e^{-\frac{u_0-u}{\beta}}+\mathcal{O}\left(\delta\right)^2 \ . 
\end{equation}
%
%We have assigned $\delta$ to be of $\mathcal{O}\left(e^{-\frac{u_0-u}{\beta}}\right)$.
Where, in the above, $\delta$ is a bookkeeping device introduced by hand to keep track of the perturbative expansion. In the end, we will set $\delta=1$.
Now, we will do the same analysis as in \S\ref{subsec:escaping} assuming the mirror profile \eqref{eq:pert_kink}.

%\subsubsection*{Conformal cross ratio}
In $(\tilde{u},\tilde{v})$ coordinates (see fig. \ref{fig:kink}), we take the bulk point to be $P\equiv(\tilde{u},\tilde{v})$, the reflected point would be $P_r\equiv(\tilde{v},\tilde{u})$. Let's assume $P_1$ and $P_2$ are two boundary points with coordinates $P_1\equiv (\tilde{v}_1,\tilde{v}_1)$ and $P_2\equiv (\tilde{v}_2,\tilde{v}_2)$. We will take the two boundary points $P_1$ and $P_2$ at a same point $P_1\equiv P_2$ which implies $\tilde{v}_1=\tilde{v}_2$.
%In $(\tilde{u},\tilde{v})$ coordinates (see Figure \ref{fig:kink}), let's assume $P_1$ and $P_2$ are two boundary points with coordinates $P_1\equiv (\tilde{v}_1,\tilde{v}_1)$ and $P_2\equiv (\tilde{v}_2,\tilde{v}_2)$. Let's assume $P$ to be a point in the bulk with coordinates $P\equiv(\tilde{u},\tilde{v})$, the reflected point $P_r$ will have coordinates $P_r\equiv(\tilde{v},\tilde{u})$.
The cross ratio is defined as:
\begin{equation}\label{eq:cross_ratio2}
\begin{split}
\eta&=\frac{(\tilde{u}+i\,\tilde{\epsilon}_0-\tilde{v}-i\,\tilde{\epsilon}_0)(\tilde{v}_1+i\,\tilde{\epsilon}_1-\tilde{v}_2-i\,\tilde{\epsilon}_2)}{(\tilde{u}+i\,\tilde{\epsilon}_0-\tilde{v}_1-i\,\tilde{\epsilon}_1)(\tilde{v}+i\,\tilde{\epsilon}_0-\tilde{v}_2-i\,\tilde{\epsilon}_2)}\\
&=\frac{(\tilde{u}-\tilde{v})i\,\tilde{\epsilon}_{12}}{(\tilde{u}-\tilde{v}_1+i\,\tilde{\epsilon}_{01})(\tilde{v}-\tilde{v}_1+i\,\tilde{\epsilon}_{02})}\\
&=\frac{\big(p(u)-v\big)i\,\tilde{\epsilon}_{12}}{\big(p(u)-v_1+i\,\tilde{\epsilon}_{01}\big)\big(v-v_1+i\,\tilde{\epsilon}_{02})}\\
&=i\,\tilde{\epsilon}_{12}\frac{-\beta\log\left(1+e^{-\frac{u}{\beta}}\right)+\delta\,\beta e^{-\frac{u_0-u}{\beta}}-v}{\left[-\beta\log\left(1+e^{-\frac{u}{\beta}}\right)+\delta\,\beta e^{-\frac{u_0-u}{\beta}}-v_1+i\,\tilde{\epsilon}_{01}\right]\left[v-v_1+i\,\tilde{\epsilon}_{02}\right]}\\
&=i\,\tilde{\epsilon}_{12}\frac{-\beta\log\left(1+e^{-\frac{u}{\beta}}\right)-v}{\left[-\beta\log\left(1+e^{-\frac{u}{\beta}}\right)-v_1+i\,\tilde{\epsilon}_{01}\right]\left[v-v_1+i\,\tilde{\epsilon}_{02}\right]}\\
&+i\delta\,\tilde{\epsilon}_{12}\frac{\beta e^{-\frac{u_0-u}{\beta}}}{\left[-\beta\log\left(1+e^{-\frac{u}{\beta}}\right)-v_1+i\tilde{\epsilon}_{01}\right]\left[v-v_1+i\tilde{\epsilon}_{02}\right]}\left[1-\frac{-\beta\log\left(1+e^{-\frac{u}{\beta}}\right)-v}{-\beta\log\left(1+e^{-\frac{u}{\beta}}\right)-v_1+i\tilde{\epsilon}_{01}}\right]+\mathcal{O}(\delta)^2\,.\\
%&-i\delta\,\epsilon_{12}\frac{-\beta\log\left(1+e^{-\frac{u+i\epsilon_0}{\beta}}\right)-v-i\epsilon_0}{\left[-\beta\log\left(1+e^{-\frac{u+i\epsilon_0}{\beta}}\right)-v_1-i\epsilon_1\right]\left[v-v_2+i\epsilon_{02}\right]}\frac{\beta e^{-\frac{u_0-u-i\epsilon_0}{\beta}}}{\left[-\beta\log\left(1+e^{-\frac{u+i\epsilon_0}{\beta}}\right)-v_1-i\epsilon_1\right]}+\mathcal{O}(\delta)^2\\
\end{split}
\end{equation}
In the analysis of kink mirror in \S\ref{subsec:kink}, where we kept $u_0$ to be fixed $\mathcal{O}(1)$ number it was not possible to set $v=0$ and then take $u\rightarrow \infty$ for the bulk operator because the kink mirror profile crosses the $u$-axis at $u_0/2$. In this section, we are always keeping $u_0>>u$. If we take $u_0>2u$, then it is possible to set $v=0$ and then take $u\rightarrow \infty$ for the bulk operator.\\\\
For example, let us now consider $v=0, u\rightarrow\infty$. In this limit,
\begin{equation}\label{eq:etaexpan}
\begin{split}
\eta&=-i\,\tilde{\epsilon}_{12}\,\frac{\beta\, e^{-u/\beta}}{v_1^2}+i\delta\,\tilde{\epsilon}_{12}\frac{\beta e^{-\frac{u_0-u}{\beta}}}{v_1^2}\left[1-\frac{\beta e^{-\frac{u}{\beta}}}{v_1}\right]+\mathcal{O}(\delta)^2 \\
&=-i\,\tilde{\epsilon}_{12}\,\frac{\beta\, e^{-u/\beta}}{v_1^2}+i\delta\,\tilde{\epsilon}_{12}\frac{\beta e^{-\frac{u_0-u}{\beta}}}{v_1^2}-i\delta\,\tilde{\epsilon}_{12}\frac{\beta e^{-\frac{u_0-u}{\beta}}}{v_1^2}\frac{\beta e^{-\frac{u}{\beta}}}{v_1}+\mathcal{O}(\delta)^2\\
&=-i\,\tilde{\epsilon}_{12}\,\frac{\beta\, e^{-u/\beta}}{v_1^2}+i\delta\,\tilde{\epsilon}_{12}\frac{\beta e^{-\frac{u_0-u}{\beta}}}{v_1^2}+\mathcal{O}(\delta)^2\,.
\end{split}
\end{equation}
Going from second line to third line, we have neglected the last term which is exponentially suppressed with respect to the second term. The above perturbative expansion will break down when
\begin{equation}
\begin{split}
&i\,\tilde{\epsilon}_{12}\,\frac{\beta\, e^{-u/\beta}}{v_1^2}\approx i\,\tilde{\epsilon}_{12}\frac{\beta e^{-\frac{u_0-u}{\beta}}}{v_1^2}\\
\Rightarrow~& u\approx u_0/2\,.
\end{split}
\end{equation}
%
%Let us redefine $u_0=\gamma \, u$ where, $\gamma>2$.
%
%\begin{equation}
%\eta=-i\,\tilde{\epsilon}_{12}\,\frac{\beta\, e^{-u\left(\frac{1}{\beta}\right)}}{v_1^2}+i\delta\,\tilde{\epsilon}_{12}\frac{\beta\,e^{-u\left(\frac{\gamma-1}{\beta}\right)}}{v_1^2}-i\delta\,\tilde{\epsilon}_{12}\frac{\beta^2\,e^{-u\left(\frac{\gamma}{\beta}\right)}}{v_1^3}+\mathcal{O}(\delta)^2 \ . 
%\end{equation}
%
The leading order term in \eqref{eq:etaexpan} is non-zero, therefore $\mathcal{O}(\delta)$ and higher order terms would not be able to change the behaviour of $\eta$. In other words, $\eta$ goes to zero from the opposite direction in the limit $u\rightarrow 0$ and $u\rightarrow\infty$, and for some intermediate point $u=u_1$ in the out-of-time-order case it crosses the real axis after $\eta=1$ branch point, whereas for time-ordered case, it crosses the real axis before the $\eta=1$ branch point.

Now, for a large-$c$ CFT, using the Virasoro Identity conformal block, we obtain:
\begin{equation}
\begin{split}
\mathcal{F}_0(\eta)
&=\left(\frac{1}{1-\frac{24\pi i h_w}{c\eta}}\right)^{2h_v}\\
&=\left(\frac{1}{1+\frac{24\pi h_w\,v_1^2}{c\,\tilde{\epsilon}_{12}\,\beta}e^{\frac{u}{\beta}}\left(1-\delta e^{-\frac{u_0-2u}{\beta}} \right)^{-1}}\right)^{2h_v}\,.
\end{split}
\end{equation}
Defining $\Lambda\equiv\frac{24\pi h_w\,v_1^2}{c\,\tilde{\epsilon}_{12}\,\beta}e^{\frac{u}{\beta}}$, we obtain:
\begin{equation}
\begin{split}
\mathcal{F}_0(\eta)
&=\left(\frac{1}{1+\Lambda\left(1+\delta e^{-\frac{u_0-2u}{\beta}}\right)}\right)^{2h_v}\\
%&=\left(\frac{1}{1+\Lambda+\Lambda\delta e^{-\frac{u_0-2u}{\beta}}}\right)^{2h_v}\\
&=\left(\frac{1}{1+\Lambda}\right)^{2h_v}\left(\frac{1}{1+\delta\frac{\Lambda}{1+\Lambda}e^{-\frac{u_0-2u}{\beta}}}\right)^{2h_v}\\
%&=\left(\frac{1}{1+\Lambda}\right)^{2h_v}\left({1-\delta\,2h_v\frac{\Lambda}{1+\Lambda}e^{-\frac{u_0-2u}{\beta}}}\right)\\
&\sim\left(\frac{1}{1+\Lambda}\right)^{2h_v}-\delta\,2h_v\left(\frac{\Lambda}{1+\Lambda}\right)\left(\frac{1}{1+\Lambda}\right)^{2h_v}e^{-\frac{u_0-2u}{\beta}} \ .
\end{split}
\end{equation}
%
%{\bf Not sure if we need the detailed formulae above.} 
It is straightforward to observe that the exponential growth in $u$ is drastically cut-off by $u_0$. In the limit $u_0 \gg u$, the exponential function has a marked existence, however for $u_0 \sim {\cal O}(1)$, this exponential growth is terminated.

Physically, this offers an intuitive understanding. In the limit when $u_0 \gg u$, the kink mirror profiles asymptotes to the escaping mirror profile, since the location of the kink is set by the parameter $u_0$. Pictorially, this is easily seen by sliding $u_0$ in fig. \ref{fig:kink} further and further away to obtain the fig. \ref{fig:escaping}. In this limit, the $b$-OTOC displays an exponential growth and therefore a scrambling dynamics. Once $u_0$ slides back into the picture, the range of $u$ get appropriately squeezed and for $u_0 \sim {\cal O}(1)$, the exponential growth yields to a power-law behaviour. Correspondingly, no scrambling is seen. Note that, it is in the latter limit where a Page curve is obtained \cite{Akal:2020twv,Akal:2021foz}. 

%{\bf we need to clearly spell out the physics of the calculations below}

%In the next section, we discuss another example of a model with a tunable parameter, such that at large values of the parameter, the system develops scrambling dynamics.
\subsection{Generalized escaping mirror}\label{gem}
We now consider another example of a model with a tunable parameter, such that at large values of the parameter, the system develops scrambling dynamics. In this case, the profile takes the form:
\begin{equation}\label{gemp}
p(u)=-\beta \log \left(1+e^{-\frac{u}{\beta}}\right)-\beta\left[\log\left(e^{\frac{u_0}{\beta}}+e^{\frac{u}{\beta}}\right)\right]^{-n}\,,
\end{equation}
which is similar to Type-B mirrors considered in \cite{Akal:2022qei}. Notice that in the asymptotic limit of $n$, the above profile reduces to that of the escaping mirror. 

%At any finite $n$, in the $u\rightarrow \infty$ limit, we can easily compute asymptoti.
%\begin{equation}\label{eq:asym_genescape}
%\begin{split}
%p(u)&=-\beta \log \left(1+e^{-\frac{u}{\beta}}\right)-\beta\left[\log\left(e^{\frac{u}{\beta}}\left(1+e^{\frac{u_0}{\beta}}e^{-\frac{u}{\beta}}\right)\right)\right]^{-n}\\
%&=-\beta \log \left(1+e^{-\frac{u}{\beta}}\right)-\beta\left[\frac{u}{\beta}+\log\left(1+e^{\frac{u_0}{\beta}}e^{-\frac{u}{\beta}}\right)\right]^{-n}\\
%&=-\beta \log \left(1+e^{-\frac{u}{\beta}}\right)-\beta\left(\frac{\beta}{u}\right)^n\left[1+\frac{\beta}{u}\log\left(1+e^{\frac{u_0}{\beta}}e^{-\frac{u}{\beta}}\right)\right]^{-%n}\\
%\end{split}
%\end{equation}
%As $u\rightarrow \infty$ limit,
%\begin{equation}
%\begin{split}
%p(u)&=-\beta\, e^{-\frac{u}{\beta}}-\beta\left(\frac{\beta}{u}\right)^n\left[1+\frac{\beta}{u}e^{\frac{u_0}{\beta}}e^{-\frac{u}{\beta}}+\cdots\right]^{-n}\\
%&=-\beta\, e^{-\frac{u}{\beta}}-\beta\left(\frac{\beta}{u}\right)^n\left[1-n\frac{\beta}{u}e^{\frac{u_0}{\beta}}e^{-\frac{u}{\beta}}+\cdots\right]\\
%&=-\frac{\beta^{n+1}}{u^n}+\mathcal{O}\left(e^{-\frac{u}{\beta}}\right)
%\end{split}
%\end{equation}
To analyze the b-OTOC, we Consider a similar set-up as discussed in \S\ref{subsec:escaping}.  From eqn. \eqref{eq:etaescaping} we get conformal cross ratio
\begin{equation}
\begin{split}
\eta&=\frac{(\tilde{u}-\tilde{v})i\,\tilde{\epsilon}_{12}}{(\tilde{u}-\tilde{v}_1+i\,\tilde{\epsilon}_{01})(\tilde{v}-\tilde{v}_1+i\,\tilde{\epsilon}_{02})}\\
&=\frac{\big(p(u)-v\big)i\,\tilde{\epsilon}_{12}}{\big(p(u)-v_1+i\,\tilde{\epsilon}_{01}\big)\big(v-v_1+i\,\tilde{\epsilon}_{02})}\\
&=i\,\tilde{\epsilon}_{12}\,\frac{-\beta\log\left(1+e^{-\frac{u}{\beta}}\right)-\beta\left[\log\left(e^{\frac{u_0}{\beta}}+e^{\frac{u}{\beta}}\right)\right]^{-n}-v}{\left[-\beta\log\left(1+e^{-\frac{u}{\beta}}\right)-\beta\left[\log\left(e^{\frac{u_0}{\beta}}+e^{\frac{u}{\beta}}\right)\right]^{-n}-v_1+i\,\tilde{\epsilon}_{01}\right]\left[v-v_1+i\,\tilde{\epsilon}_{02}\right]}\,.
\end{split}
\end{equation}
\subsection*{Case-1: $v=0, u\rightarrow\infty$}

In the asymptotic $u$ limit, the profile \eqref{gemp} takes the form: 
\begin{equation}
p(u) \sim -\frac{\beta^{n+1}}{u^n}+\mathcal{O}\left(e^{-\frac{u}{\beta}}\right)\,.
%\begin{split}
%p(u)&=-\beta\, e^{-\frac{u}{\beta}}-\beta\left(\frac{\beta}{u}\right)^n\left[1+\frac{\beta}{u}e^{\frac{u_0}{\beta}}e^{-\frac{u}{\beta}}+\cdots\right]^{-n}\\
%&=-\beta\, e^{-\frac{u}{\beta}}-\beta\left(\frac{\beta}{u}\right)^n\left[1-n\frac{\beta}{u}e^{\frac{u_0}{\beta}}e^{-\frac{u}{\beta}}+\cdots\right]\\
%&=-\frac{\beta^{n+1}}{u^n}+\mathcal{O}\left(e^{-\frac{u}{\beta}}\right)
%\end{split}
\end{equation}
Therefore in this limit, the cross-ratio takes the form:
%eqn. \eqref{eq:asym_genescape},
\begin{equation}
\eta=i\tilde{\epsilon}_{12}\frac{-\beta^{n+1}}{u^n\,v_1^2}\,.
\end{equation}

\subsection*{Case-2: $v=0, u=0$}
\begin{equation}
\eta=i\tilde{\epsilon}_{12}\frac{\beta \log 2+\beta\left[\log\left(1+e^{\frac{u_0}{\beta}}\right)\right]^{-n}}{\left(\beta\log 2+\beta\left[\log\left(1+e^{\frac{u_0}{\beta}}\right)\right]^{-n}+v_1\right)(-v_1)}\,.
\end{equation}
$-v_1>0$, and let's assume $\beta\log 2+\beta\left[\log\left(1+e^{\frac{u_0}{\beta}}\right)\right]^{-n}+v_1>0$. The above assumptions imply that the points $P_1$ and $P_2$ have to lie in the quadrant $(u>0,v<0)$ in the $(u,v)$ plane. So we see that $\eta$ approaches $0$ from the opposite directions for Case-1 and Case-2.

\subsection*{Case-3: $v=0, u=u_1=p^{-1}(v_1)$}
\begin{equation}
\begin{split}
\eta&=i\,\tilde{\epsilon}_{12}\frac{-\beta\log\left(1+e^{-\frac{u_1}{\beta}}\right)-\beta\left[\log\left(e^{\frac{u_0}{\beta}}+e^{\frac{u_1}{\beta}}\right)\right]^{-n}}{\left[-\beta\log\left(1+e^{-\frac{u_1}{\beta}}\right)-\beta\left[\log\left(e^{\frac{u_0}{\beta}}+e^{\frac{u_1}{\beta}}\right)\right]^{-n}-v_1+i\,\tilde{\epsilon}_{01}\right][-v_1]}\\
&=-\frac{\tilde{\epsilon}_{12}}{\tilde{\epsilon}_{01}}=1+\frac{\tilde{\epsilon}_{20}}{\tilde{\epsilon}_{01}}\,.
\end{split}
\end{equation}
Cross-ratio shows similar behaviour as that of the escaping mirror. From \eqref{eq:vac_vir}, we can very easily see that as $u\rightarrow\infty$
\begin{equation}
\begin{split}
\mathcal{F}_0(\eta)
&=\left(\frac{1}{1-\frac{24\pi i h_w}{c\eta}}\right)^{2h_v}\\
&=\left(\frac{1}{1+\frac{24\pi  h_w u^n v_1^2}{c\,\tilde{\epsilon}_{12}\beta^{n+1}}}\right)^{2h_v}\\
&\approx 1-\frac{48\pi h_v h_w v_1^2}{c\tilde{\epsilon}_{12}\beta^{n+1}}u^n+\cdots\,.
\end{split}
\end{equation}
For all finite $n\geq 0$ it shows polynomial behaviour.

\section{Two-point function}\label{2pt}
It is well-known that the thermal two-point function decays exponentially at late times. From the analysis of the previous section, we have seen that the escaping mirror profile effectively induces a temperature set by the parameter $\beta$ though this is not the case for the kink mirror profile. In this section, we will compute two-point function with one operator on the boundary and another in the bulk subjected to the escaping mirror and kink mirror boundary conditions.

\subsection{Escaping mirror}
We compute two-point function with one operator $V$ on the boundary at the point $P_0\equiv(\tilde{v}_0,\tilde{v}_0)$ and another operator $W$ in the bulk at the point $P\equiv (\tilde{u},\tilde{v})$ subjected to the escaping mirror boundary condition. Using Cardy's {\it doubling trick}, this two-point function will be the same as three-point function without boundary with one extra operator $W$ at the reflected point  $P_r\equiv (\tilde{v},\tilde{u})$. Therefore, we get
\begin{equation}
\begin{split}
\langle W(u,v) V(u_0,v_0)\rangle&\sim\frac{\gamma}{(\tilde{u}-\tilde{v}_0)^{h_w+h_v-h_w}(\tilde{v}-\tilde{u})^{2h_w-h_v}(\tilde{v}-\tilde{v}_0)^{h_w+h_v-h_w}}\\
&=\frac{\gamma}{[p(u)-v_0]^{h_v}[v-p(u)]^{2h_w-h_v}[v-v_0]^{h_v}}\,.
\end{split}
\end{equation}
Where $\gamma$ is some constant. Now, if for our bulk operator we set $v=0$ and then take $u\rightarrow \infty$, we get
\begin{equation}
\begin{split}
\langle W(u,v) V(u_0,v_0)\rangle&\sim \frac{\gamma}{v_0^{2h_v}\left[\beta\log\Big(1+e^{-\frac{u}{\beta}}\Big)\right]^{2h_w-h_v}}\\
&=\frac{\gamma}{v_0^{2h_v} \beta^{2h_w-h_v}}e^{-\frac{u}{\beta}(h_v-2h_w)}\,.
\end{split}
\end{equation}
It decays exponentially fast as $u\rightarrow \infty$.

\subsection{Kink mirror}
Similarly, we can compute two-point function in kink mirror profile \eqref{eq:kinkmirror}. We take the operator $V$ on the boundary at a point $P_0\equiv (\tilde{v}_0,\tilde{v}_0)$ and another operator $W$ in the bulk at a point $P\equiv (\tilde{u},\tilde{v})$. The reflected point is $P_r\equiv (\tilde{v},\tilde{u})$. The two-point function is given by
\begin{equation}
\begin{split}
\langle W(u,v) V(u_0,v_0)\rangle&\sim\frac{\gamma}{(\tilde{u}-\tilde{v}_0)^{h_w+h_v-h_w}(\tilde{v}-\tilde{u})^{2h_w-h_v}(\tilde{v}-\tilde{v}_0)^{h_w+h_v-h_w}}\\
&=\frac{\gamma}{[p(u)-v_0]^{h_v}[v-p(u)]^{2h_w-h_v}[v-v_0]^{h_v}}\,.
\end{split}
\end{equation}
Now, if for our bulk operator we set $v=0$ and then take $u=\frac{u_0}{2}-\alpha$, $\alpha\rightarrow 0$, we get
\begin{equation}
\begin{split}
\langle W(u,v) V(u_0,v_0)\rangle&\sim \frac{\gamma}{v_0^{2h_v}\left(\frac{2\alpha}{1+e^{\frac{u_0}{2\beta}}}\right)^{2h_w-h_v}}\\
&=\frac{\gamma}{v_0^{2h_v}}\left(\frac{2\alpha}{1+e^{\frac{u_0}{2\beta}}}\right)^{h_v-2h_w}\,.
\end{split}
\end{equation}
It decays polynomially as $\alpha\rightarrow 0$.

%\section{Ising model and orbifold CFTs}
%For ising model (see appendix B of %\cite{Roberts:2014ifa})
%\begin{equation}
%f_{\sigma\epsilon}(\eta,\bar{\eta})= \left(\frac{2-\eta}{2\sqrt{1-\eta}}\right)\left(\frac{2-\bar{\eta}}{2\sqrt{1-\bar{\eta}}}\right)
%\end{equation}
%As we take $t\rightarrow\infty$, both $\eta,\bar{\eta}\rightarrow 0$. But, $\eta$ crosses the brunch point at $\eta=1$.
%\begin{equation}
%\begin{split}
%f_{\sigma\epsilon}(\eta,\bar{\eta})&=\frac{2-\eta}{2(1-\frac{1}{2}\eta)e^{-i\pi}}\,\frac{2-\bar\eta}{2(1-\frac{1}{2}\bar\eta)}\\
%&=-1
%\end{split}
%\end{equation}
%Ising model does not show chaotic behaviour with moving mirror boundary condition.
%
%\subsection*{Orbifold CFT $(T^2)^2/Z_2$}
%For orbifold CFT $(T^2)^2/Z_2$ (see eq.(3.29) of \cite{Caputa:2017rkm})
%\begin{equation}
%\begin{split}
%F_2(\eta,\bar{\eta})&\simeq \frac{i\pi}{2\log\left(\frac{\eta}{16}\right)}\\
%&=\frac{i\pi}{2\log\left(-i\frac{\tilde{\epsilon}_{12}\beta}{16 v_1^2}e^{-\frac{u}{\beta}}\right)}
%\end{split}
%\end{equation}

\section{Discussions}\label{discussion}

In this article, we have explored the physics of scrambling in a wide class of moving mirror models and within a two-dimensional CFT. We have observed that scrambling, defined in terms of OTOCs, exhibits a class of desired properties ranging from (maximally) chaotic dynamics to a unitary evolution of black holes, within the framework of moving mirror models. We have also observed that most mirror profiles yield a power-law behaviour of OTOC and therefore do not exhibit an exponential-in-time scrambling physics. The power of the polynomial-in-time behaviour is mostly governed by the conformal dimension of the heavy operator of the Virasoro identity block. It is therefore possible to approximate an exponential behaviour for a range of reasonably large-time dynamics by such a power-law. Nonetheless, such a scrambling quickly yields for a unitary physics.\footnote{Roughly, we can always write:
\begin{eqnarray}
t^{\Delta} = e^{\Delta \log t} = e^{\Delta t \left( \frac{\log t}{t} \right)} = e^{\Delta(t) t}\ ,
\end{eqnarray}
where $\Delta(t) = \frac{\log t}{t}$ can be identified with a time-dependent Lyapunov exponent. The function $\Delta(t)$ varies reasonably slowly near $t \approx e$ and therefore the exponential behaviour can be interpreted as a Lyapunov growth. 
}
Note that, this transient exponential behaviour is associated with a Lyapunov exponent which can be arbitrarily large in principle. This is in no contradiction with the chaos-bound of \cite{Maldacena:2015waa} since the latter strictly applies for time-independent thermal state.

It is worthwhile to emphasize again the escaping mirror and the kink mirror models. While the former captures physics of an eternal black hole and correspondingly displays a fast scrambling, maximally chaotic behaviour, the latter demonstrates how the exponential growth yields to a power law behaviour near the Page time and therefore becomes a slow scrambler towards the later part of the dynamics. Qualitatively, it is connected to unitarity of the dynamics, since near the Page time information begins being retrieved and thus a fast scrambling behaviour is expected to disappear. Furthermore, this provides a simple and tractable model where an interpolation from fast scrambling to a slow scrambling dynamics can be realized as a function of a single parameter.

Our results are in qualitative agreement with the physics discussed in \cite{Cotler:2022weg}, where it was shown that an escaping mirror profile gives rise to a non-unitary dynamics at the future infinity. It was argued that this is arising from throwing away details beyond a UV cut-off which is essential for the computation of entanglement entropy. At very late times, these UV-modes get Doppler shifted back and contribute to the entanglement entropy which should yield for a unitary dynamics. This physics has a very close parallel to the physics of scrambling in terms of the kink-mirror, as a function of the location of the kink. In the escaping mirror limit, we observe a perfect exponential growth with a maximal Lyapunov exponent which yields for a power-law and therefore unitary dynamics when the kink is located at a finite distance. In a qualitative sense, the escaping mirror limit allows for degrees of freedom to escape from the system which are brought back when the kink appears at a finite distance.

Note further that, in the strict ``escaping mirror" limit, the density matrix corresponds to a mixed thermal density matrix and an emergent KMS-condition is expected. This happens in the strict $u_0 \to \infty$ limit. For any finite $u_0$, a KMS-condition can only be a transient approximate relation. This is reminiscent of a type I$_\infty$ von Neumann algebra approximating a type III von Neumann algebra. In recent times, there has been a surge of activities to better understand aspects of operator algebra and factorization, including possible transitions between them, in quantum mechanical as well as quantum field theoretic models, see {\it e.g.}~\cite{Witten:2021jzq, Witten:2021unn, Witten:2023qsv, Banerjee:2023eew, Banerjee:2023liw, Basteiro:2024cuh}. It will be extremely interesting to explicitly realize such transitions/crossover of algebras within a simple model like CFT. We hope to address this aspect in near future.

There are several other future directions and we will briefly discuss some of them now. First, by now there is a standard literature on Holographic descriptions of such mirror models as well as models with defects and interfaces in the CFT. It will be interesting to realize the physics from a purely bulk geometric perspective. This is not just for reassurance, several technical questions often have an intriguing geometric understanding in the Holographic dual, {\it e.g.}~global and local quenches (see {\it e.g.}~\cite{Das:2011nk} for a review), OTOCs and shockwave geometries (see {\it e.g.}~\cite{Shenker:2013pqa, Shenker:2014cwa}) and many more. 

The Holographic dual of the kink mirror is particularly intriguing in this respect. It is well-known that the exponential growth of an OTOC originates from the blue-shift near the horizon and from a corresponding shockwave geometry. The kink mirror model somehow ensures that this blue-shift will disappear as a result of the dynamical evaporation of the black hole. It will be very interesting to understand this effect in explicit details. In a precise sense, all essential ingredients are already available, and we hope to address this issue in near future.

Generally speaking, the presence of boundaries in a CFT is realized by inserting branes and hypersurfaces in the bulk Holographic dual. Sometimes, these branes are the so-called End-of-World (EOW) branes. There is a broader question on how the presence of such branes may affect the dynamics of {\it e.g.}~a probe field in a given asymptotically AdS-geometry. Recent studies in \cite{Das:2022evy, Das:2023ulz, Das:2023xjr, Krishnan:2023jqn, Burman:2023kko, Banerjee:2024dpl} suggest that a non-generic form of chaos may appear as a function of the location of a geometric brane. It is, however, not understood how a corresponding OTOC may behave in these cases which will be a very interesting aspect to address in future.

It is worthwhile to pause at this point and reflect back on the possibilities that the moving mirror models can capture in a broad physics sense. It is clear that this framework allows us to explicitly calculate and catalogue OTOC behaviours of two-dimensional CFTs in these backgrounds and therefore it will be highly relevant to understand in detail the possible ramifications of such explicit results on the physical aspects that the moving mirror models can represent.

\section*{Acknowledgements}
We thank Souvik Banerjee, Suchetan Das, Anirban Dinda, Johanna Erdmenger, Jani Kastikainen, Dominik Neuenfeld, and Somnath Porey for useful discussions related to this work. PB is supported by the grant CRG/2021/004539. BE acknowledges the support provided by the grant CRG/2021/004539. AK is partially supported by CEFIPRA 6304 - 3, DAE-BRNS 58/14/12/2021-BRNS and CRG/2021/004539 of Govt. of India. AK further acknowledges the support of Humboldt Research Fellowship for Experienced Researchers by the Alexander von Humboldt Foundation and for the hospitality of the Theoretical Physics III, Department of Physics and Astronomy, University of Wurzburg during the course of this work.
%BR is supported by the Senior Research Fellowship(SRF) funded by the University Grant Commission(UGC) under the CSIR-UGC NET Fellowship.

\appendix

\section{Analytic continuation in $(u,v)$ coordinates}\label{app:analytic}
Different operators in OTOCs get ordered according to their Lorentzian continuation. This appendix will explain the analytic continuation in $(u,v)$ coordinates. We
take our bulk operator $W$ with scaling dimension $h_w$ at the point $P$ and the boundary operator $V$ with scaling dimension $h_v$ at points $P_1$ and $P_2$.
\begin{equation}
\begin{split}
P_1&\equiv (v_1+i\epsilon_1, v_1+i\epsilon_1)\,,\qquad P_2\equiv (v_1+i\epsilon_2, v_1+i\epsilon_2)\,,\\
P&\equiv (u+i\epsilon_0, v+i\epsilon_0)\,, \quad\quad~~ P_r\equiv (v+i\epsilon_0, u+i\epsilon_0)\,.
\end{split}
\end{equation}
We will take the two boundary points $P_1$ and $P_2$ at the same point $P_1\equiv P_2$ which implies $v_1=v_2$. The conformal cross ratio is given by
\begin{equation}
\begin{split}
\eta&=\frac{(\tilde{u}-\tilde{v})(\tilde{v}_1-\tilde{v}_2)}{(\tilde{u}-\tilde{v}_1)(\tilde{v}-\tilde{v}_2)}\\
&=\frac{\Big[p(u+i\epsilon_0)-v-i\epsilon_0\Big]\Big[v_1+i\epsilon_1-v_1-i\epsilon_2\Big]}{\Big[p(u+i\epsilon_0)-v_1-i\epsilon_1\Big]\Big[v+i\epsilon_0-v_1-i\epsilon_2\Big]}\\
&=i\epsilon_{12}\frac{-\beta\log\left(1+e^{-\frac{u+i\epsilon_0}{\beta}}\right)-v-i\epsilon_0}{\left[-\beta\log\left(1+e^{-\frac{u+i\epsilon_0}{\beta}}\right)-v_1-i\epsilon_1\right]\Big[v-v_1+i\epsilon_{02}\Big]}\,.
\end{split}
\end{equation}
\subsection*{Case-1: $v=0, u\rightarrow\infty$}
\begin{equation}
\eta=-\frac{i\epsilon_{12}\,\beta\, e^{-u/\beta}}{v_1^2}\,.
\end{equation}

\subsection*{Case-2: $v=0, u=0$}
\begin{equation}
\eta=\frac{i\epsilon_{12}\, \beta \log 2}{(\beta\log 2+v_1)(-v_1)}\,.
\end{equation}
$-v_1>0$, and let's assume $\beta\log 2+v_1>0$. The above assumption implies that the points $P_1$ and $P_2$ have to lie in the quadrant $(u>0,v<0)$ in the $(u,v)$ plane. So we see that $\eta$ approaches $0$ from the opposite directions for Case-1 and Case-2.

%\subsection*{Case-3: Null $v=v_1$}
%\begin{equation}
%\eta=\frac{\epsilon_{12}}{\epsilon_{02}}=\frac{\epsilon_{21}}{\epsilon_{20}}=1+\frac{\epsilon_{01}}{\epsilon_{20}}
%\end{equation}

\subsection*{Case-3: $v=0, u=u_1=p^{-1}(v_1)$}

\begin{equation}
\begin{split}
\eta&=i\epsilon_{12}\frac{-\beta\log\left(1+e^{-\frac{u_1+i\epsilon_0}{\beta}}\right)}{\left[-\beta\log\left(1+e^{-\frac{u_1+i\epsilon_0}{\beta}}\right)-v_1-i\epsilon_1\right][-v_1]}\\
&=-\frac{i\epsilon_{12}}{-\beta \log\left[\left(1+e^{-\frac{u_1}{\beta}}\right)\left(1-\frac{i\epsilon_0}{\beta}\frac{e^{-\frac{u_1}{\beta}}}{1+e^{-\frac{u_1}{\beta}}}\right)\right]-v_1-i\epsilon_1}\\
&=-\frac{i\epsilon_{12}}{i\epsilon_0 \frac{e^{-\frac{u_1}{\beta}}}{1+e^{-\frac{u_1}{\beta}}}-i\epsilon_1}\,.
\end{split}
\end{equation}
Let's define $\epsilon_{\tilde{0}}=\epsilon_0 \frac{e^{-\frac{u_1}{\beta}}}{1+e^{-\frac{u_1}{\beta}}}$. Under this redefinition the cross-ratio becomes
\begin{equation}
\begin{split}
\eta&=-\frac{\epsilon_{12}}{\epsilon_{\tilde{0}1}}=1+\frac{\epsilon_{2\tilde{0}}}{\epsilon_{\tilde{0}1}}\,.
\end{split}
\end{equation}
The cross ratio $\eta$ goes to zero from the opposite direction in the limit $u\rightarrow 0$ and $u\rightarrow\infty$. It crosses the real axis after $\eta=1$ branch point
if we choose $\epsilon_2>\epsilon_{\tilde{0}}>\epsilon_1$ (or, $\epsilon_1>\epsilon_{\tilde{0}}>\epsilon_2$). We can choose $\epsilon$'s in such a way that $\epsilon_2>\epsilon_{0}>\epsilon_1\Rightarrow \epsilon_2>\epsilon_{\tilde{0}}>\epsilon_1$ but there are choices for which $\epsilon_2>\epsilon_{0}>\epsilon_1\nRightarrow \epsilon_2>\epsilon_{\tilde{0}}>\epsilon_1$. In other words, in $(u,v)$ plane a subset of OTOCs would show the above behaviour. 

In all the previous computations in the literature, OTOCs are always determined by the analytic continuation in the static case. Now, in our case, static mirror profile is conformally related to the moving mirror profile. The above ordering issue is due to the fact that different operators that are ordered in a certain fashion in static case can become differently ordered because of conformal transformation.

\section{Other mirror profiles}\label{app: other}
\subsection{Collapsing null shell}
The mirror profile \cite{PhysRevD.36.2327,Akal:2021foz}
\begin{equation}\label{eq:collapsemirror}
v=p(u)=-\frac{1}{\kappa}e^{-\kappa u}
\end{equation}
mimics black hole that is formed from a collapsing null shell.
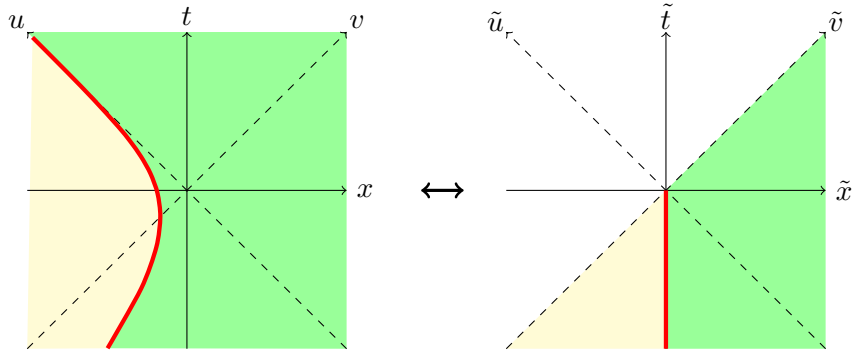
\begin{figure}[!h]
\centering
\begin{tikzpicture}[scale=0.7]
\fill[fill=green!40] (-3,3)-- plot [smooth] coordinates {(-2.901, 2.898) (-2.2555, 2.2444) (-1.765, 1.734) (-1.2910, 1.2089) (-1.0676, 0.9323) (-0.8615, 0.6384) (-0.6839, 0.3160) (-0.5532, -0.0532) (-0.5, -0.5) (-0.5743, -1.074) (-0.8591, -1.8591) (-1.4908, -2.9908)}--(3,-3)--(3,3);
\fill[fill=yellow!20,] (-3,-3)--plot [smooth] coordinates {(-2.901, 2.898) (-2.2555, 2.2444) (-1.765, 1.734) (-1.2910, 1.2089) (-1.0676, 0.9323) (-0.8615, 0.6384) (-0.6839, 0.3160) (-0.5532, -0.0532) (-0.5, -0.5) (-0.5743, -1.074) (-0.8591, -1.8591) (-1.4908, -2.9908)}--(-3,-3);
\draw [ ->] (0,-3) -- (0,3) node at (0,3.3) {$ t$};
\draw [ ->] (-3,0) -- (3,0) node[right] {$x$};
\draw [dashed, ->] (-3,-3) -- (3,3) node at (3.2,3.2) {$v$};
\draw [dashed, ->] (3,-3) -- (-3,3) node at (-3.2,3.2) {$u$};
\draw[red,ultra thick] plot [smooth] coordinates {(-2.901, 2.898) (-2.2555, 2.2444) (-1.765, 1.734) (-1.2910, 1.2089) (-1.0676, 0.9323) (-0.8615, 0.6384) (-0.6839, 0.3160) (-0.5532, -0.0532) (-0.5, -0.5) (-0.5743, -1.074) (-0.8591, -1.8591) (-1.4908, -2.9908)};
\draw[<->, very thick] (4.4,0)--(5.2,0);
\fill[fill=green!40] (9,0)--(12,3)--(12,-3)--(9,-3);
\fill[fill=yellow!20,] (9,0)--(6,-3)--(9,-3);
\draw [ ->] (9,-3) -- (9,3) node at (9,3.3) {$\tilde{t}$};
\draw [ ->] (6,0) -- (12,0) node[right] {$\tilde{x}$};
\draw [dashed, ->] (6,-3) -- (12,3) node at (12.2,3.2) {$\tilde{v}$};
\draw [dashed, ->] (12,-3) -- (6,3) node at (5.8,3.2) {$\tilde{u}$};
\draw[red,ultra thick] (9,0) -- (9,-3);
\end{tikzpicture}
\caption{Conformal mapping for moving mirror corresponding to a collapsing null shell.}
\label{fig:collapse}
\end{figure}

In the limit $u\rightarrow \infty$ the escaping mirror becomes \eqref{eq:escap_inf}
\begin{equation}
\lim_{u \to \infty} p(u) = -\beta\, e^{-u/\beta}\,.
\end{equation}
In the $u\rightarrow\infty$ limit both the escaping mirror \eqref{eq:escaping} and the collapsing mirror\eqref{eq:collapsemirror} give same behaviour. which is also evident from figure \ref{fig:escaping} and figure \ref{fig:collapse}. But, in the limit $u\rightarrow -\infty$ the escaping mirror becomes 
\begin{equation}
\lim_{u \to -\infty} p(u) = u\,,
\end{equation}
this behaviour is different for collapsing and escaping mirrors.

Considering similar set-up as discussed in \S\ref{subsec:escaping}, we can very easily compute conformal cross ratios
\subsection*{Case-1: $v=0, u\rightarrow\infty$}
\begin{equation}
\eta=-i\,\tilde{\epsilon}_{12}\frac{1}{\kappa}\frac{ e^{-\kappa u}}{v_1^2} \ .
\end{equation}

\subsection*{Case-2: $v=0, u=0$}
\begin{equation}
\eta=i\tilde{\epsilon}_{12}\frac{1/\kappa}{(1/\kappa+v_1)(-v_1)} \ .
\end{equation}
$-v_1>0$, and let's assume $1/\kappa+v_1>0$. The above assumption implies that the points $P_1$ and $P_2$ have to lie in the quadrant $(u>0,v<0)$ in the $(u,v)$ plane. So we see that $\eta$ approaches $0$ from the opposite directions for Case-1 and Case-2.

\subsection*{Case-3: $v=0, u=u_1=p^{-1}(v_1)$}
\begin{equation}
\begin{split}
\eta
&=-\frac{\tilde{\epsilon}_{12}}{\tilde{\epsilon}_{01}}=1+\frac{\tilde{\epsilon}_{20}}{\tilde{\epsilon}_{01}} \ .
\end{split}
\end{equation}
For the computation of the Virasoro identity block the $u\rightarrow \infty$ result is important and this is the same as that of the escaping mirror. This is because both the escaping mirror and the collapsing mirror behave similarly in the limit $u\rightarrow\infty$. From \eqref{eq:vac_vir}, our final result for the Virasoro identity block is given by
\begin{equation}
\begin{split}
\mathcal{F}_0(\eta)&=\left(\frac{1}{1+\frac{24\pi  h_w\,\kappa\, v_1^2}{c\, \tilde{\epsilon}_{12}}e^{\kappa\, u}}\right)^{2h_v} \approx  1-\frac{48\pi h_v h_w\,\kappa\, v_1^2}{c\, \tilde{\epsilon}_{12}}e^{\kappa\, u}+\cdots\,.
\end{split}
\end{equation}
Which again shows exponential behaviour at large $u$.

\subsection{Rindler mirror}
The mirror profile is given by \cite{Akal:2021foz}
\begin{equation}
v=p(u)=-\frac{\sigma^2}{u}\,.
\end{equation}
\begin{figure}[!h]
\centering
\begin{tikzpicture}[scale=0.7]
\fill[fill=green!40] (2.85,-2.85)-- plot [smooth] coordinates {(-3.0095, -2.8385) (-2.3827, -2.1627) (-2.125, -1.875) (-1.8167, -1.5167) (-1.6036, -1.2536) (-1.45,-1.05) (-1.3361,-0.8861) (-1.25,-0.75) (-1.1841,-0.6341) (-1.1333, -0.5333) (-1.0942,-0.4442) (-1.0643, -0.3643) (-1.0417, -0.29167) (-1.025,-0.225) (-1.0132, -0.1632) (-1.0056,-0.10556) (-1.0013,-0.0513) (-1, 0) (-1.0012,0.04881) (-1.0045,0.095455) (-1.0098,0.14022) (-1.0167,0.18333) (-1.025,0.225) (-1.0346,0.26538) (-1.0454,0.30463) (-1.0571,0.34286) (-1.0833,0.41667) (-1.1125,0.4875) (-1.1778,0.62222) (-1.25,0.75) (-1.45,1.05) (-1.6667,1.3333) (-1.8929,1.6071) (-2.125,1.875) (-2.3611,2.1389) (-2.6,2.4) (-2.8409,2.6591) (-2.9862,2.8138)}--(2.85,2.85);
\fill[fill=yellow!20,] (-3,0)--plot [smooth] coordinates {(-3.0095, -2.8385) (-2.3827, -2.1627) (-2.125, -1.875) (-1.8167, -1.5167) (-1.6036, -1.2536) (-1.45,-1.05) (-1.3361,-0.8861) (-1.25,-0.75) (-1.1841,-0.6341) (-1.1333, -0.5333) (-1.0942,-0.4442) (-1.0643, -0.3643) (-1.0417, -0.29167) (-1.025,-0.225) (-1.0132, -0.1632) (-1.0056,-0.10556) (-1.0013,-0.0513) (-1, 0) (-1.0012,0.04881) (-1.0045,0.095455) (-1.0098,0.14022) (-1.0167,0.18333) (-1.025,0.225) (-1.0346,0.26538) (-1.0454,0.30463) (-1.0571,0.34286) (-1.0833,0.41667) (-1.1125,0.4875) (-1.1778,0.62222) (-1.25,0.75) (-1.45,1.05) (-1.6667,1.3333) (-1.8929,1.6071) (-2.125,1.875) (-2.3611,2.1389) (-2.6,2.4) (-2.8409,2.6591) (-2.9862,2.8138)}--(-3,0);
\draw [ ->] (0,-3) -- (0,3) node at (0,3.3) {$ t$};
\draw [ ->] (-3,0) -- (3,0) node[right] {$x$};
\draw [dashed, ->] (-3,-3) -- (3,3) node at (3.2,3.2) {$v$};
\draw [dashed, ->] (3,-3) -- (-3,3) node at (-3.2,3.2) {$u$};
%\draw[red,ultra thick] plot [smooth] coordinates {(-3.205, -3.045) (-1.08758, -0.427576) (-1.01103, 0.148966) (-1.1312, 0.528795) (-1.31148, 0.848519) (-1.51797, 1.14203) (-1.73823, 1.42177) (-1.96661, 1.69339) (-2.20019, 1.95981) (-2.4373, 2.2227) (-2.6769, 2.4831) (-2.91834, 2.74166) (-3.16117, 2.99883)};
\draw[<->, very thick] (4.4,0)--(5.2,0);
\draw[red,ultra thick] plot [smooth] coordinates {(-3.0095, -2.8385) (-2.3827, -2.1627) (-2.125, -1.875) (-1.8167, -1.5167) (-1.6036, -1.2536) (-1.45,-1.05) (-1.3361,-0.8861) (-1.25,-0.75) (-1.1841,-0.6341) (-1.1333, -0.5333) (-1.0942,-0.4442) (-1.0643, -0.3643) (-1.0417, -0.29167) (-1.025,-0.225) (-1.0132, -0.1632) (-1.0056,-0.10556) (-1.0013,-0.0513) (-1, 0) (-1.0012,0.04881) (-1.0045,0.095455) (-1.0098,0.14022) (-1.0167,0.18333) (-1.025,0.225) (-1.0346,0.26538) (-1.0454,0.30463) (-1.0571,0.34286) (-1.0833,0.41667) (-1.1125,0.4875) (-1.1778,0.62222) (-1.25,0.75) (-1.45,1.05) (-1.6667,1.3333) (-1.8929,1.6071) (-2.125,1.875) (-2.3611,2.1389) (-2.6,2.4) (-2.8409,2.6591) (-2.9862,2.8138)};
\fill[fill=green!40] (9,0)--(12,3)--(12,-3)--(9,-3);
\fill[fill=yellow!20,] (9,0)--(6,-3)--(9,-3);
\draw [ ->] (9,-3) -- (9,3) node at (9,3.3) {$\tilde{t}$};
\draw [ ->] (6,0) -- (12,0) node[right] {$\tilde{x}$};
\draw [dashed, ->] (6,-3) -- (12,3) node at (12.2,3.2) {$\tilde{v}$};
\draw [dashed, ->] (12,-3) -- (6,3) node at (5.8,3.2) {$\tilde{u}$};
\draw[red,ultra thick] (9,0) -- (9,-3);
\end{tikzpicture}
\caption{Conformal mapping for Rindler mirror.}
\label{fig:Rindler}
\end{figure}
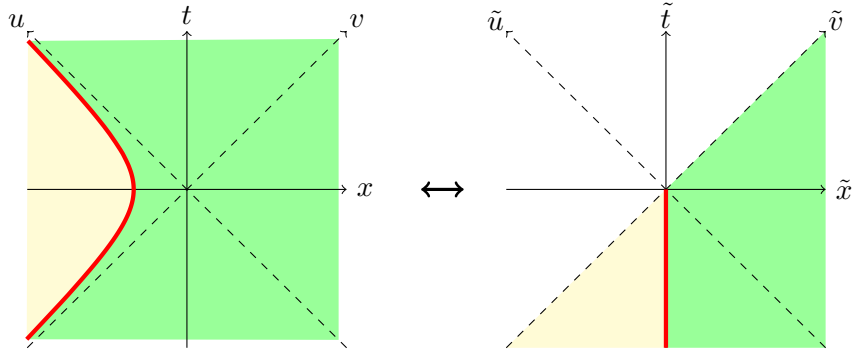
Considering similar set-up as discussed in \S\ref{subsec:escaping}, from \eqref{eq:etaescaping} we get conformal cross ratio
\begin{equation}
\begin{split}
\eta&=\frac{(\tilde{u}-\tilde{v})i\,\tilde{\epsilon}_{12}}{(\tilde{u}-\tilde{v}_1+i\,\tilde{\epsilon}_{01})(\tilde{v}-\tilde{v}_1+i\,\tilde{\epsilon}_{02})}\\
&=\frac{\big(p(u)-v\big)i\,\tilde{\epsilon}_{12}}{\big(p(u)-v_1+i\,\tilde{\epsilon}_{01}\big)\big(v-v_1+i\,\tilde{\epsilon}_{02})}\\
&=i\,\tilde{\epsilon}_{12}\,\frac{-\frac{\sigma^2}{u}-v}{\left[-\frac{\sigma^2}{u}-v_1+i\,\tilde{\epsilon}_{01}\right]\left[v-v_1+i\,\tilde{\epsilon}_{02}\right]}\,.
\end{split}
\end{equation}
\subsection*{Case-1: $v=0, u\rightarrow\infty$}
\begin{equation}
\eta=-i\,\tilde{\epsilon}_{12}\frac{ \sigma^2}{v_1^2\, u} \ .
\end{equation}

\subsection*{Case-2: $v=0, u=0$}
\begin{equation}
\eta=i\tilde{\epsilon}_{12}\frac{1}{(-v_1)} \ .
\end{equation}
$-v_1>0$. So we see that $\eta$ approaches $0$ from the opposite directions for Case-1 and Case-2.

\subsection*{Case-3: $v=0, u=u_1=p^{-1}(v_1)$}
\begin{equation}
\begin{split}
\eta
&=i\tilde{\epsilon}_{12}\frac{-\sigma^2/u_1}{[-\sigma^2/u_1-v_1+i\tilde{\epsilon}_{01}][-v_1]}\\
&=-\frac{\tilde{\epsilon}_{12}}{\tilde{\epsilon}_{01}}=1+\frac{\tilde{\epsilon}_{20}}{\tilde{\epsilon}_{01}} \ .
\end{split}
\end{equation}
The cross ratio $\eta$ goes to zero from the opposite direction in the limit $u\rightarrow 0$ and $u\rightarrow \infty$. 
For the intermediate point $u=u_1$, in the out of time order case, $\tilde{\epsilon}_2>\tilde{\epsilon}_0>\tilde{\epsilon}_1$ or, $\tilde{\epsilon}_1>\tilde{\epsilon}_0>\tilde{\epsilon}_2$, it crosses the real axis after $\eta=1$ branch point.

For a large-$c$ CFT, as before, we can again use the Identity block result. We can very easily see that as $u\rightarrow\infty$:
\begin{equation}
\begin{split}
\mathcal{F}_0(\eta)
&=\left(\frac{1}{1-\frac{24\pi i h_w}{c\eta}}\right)^{2h_v}\\
&=\left(\frac{1}{1+\frac{24\pi  h_w v_1^2}{c\,\tilde{\epsilon}_{12}\sigma^2}u}\right)^{2h_v}\\
&\approx 1-\frac{48 \pi h_v h_w v_1^2}{c\,\tilde{\epsilon}_{12}\sigma^2}u+\cdots\,.
\end{split}
\end{equation}
In this case, we get polynomial behaviour.

\subsection{Receding mirror}
In this appendix, we consider the following mirror profile \cite{Cotler:2022weg}
\begin{equation}
\begin{split}
    v=p(u)&=u,\quad u<0,\\
    &= \frac{1}{2\pi T_H}\left(1-e^{-2\pi T_H u}\right),\quad 0<u<L,\\
    &=e^{-2\pi T_H L}(u-L)+\frac{1}{2\pi T_H}\left(1-e^{-2\pi T_H L}\right),\quad u>L\,.
\end{split}
\end{equation}

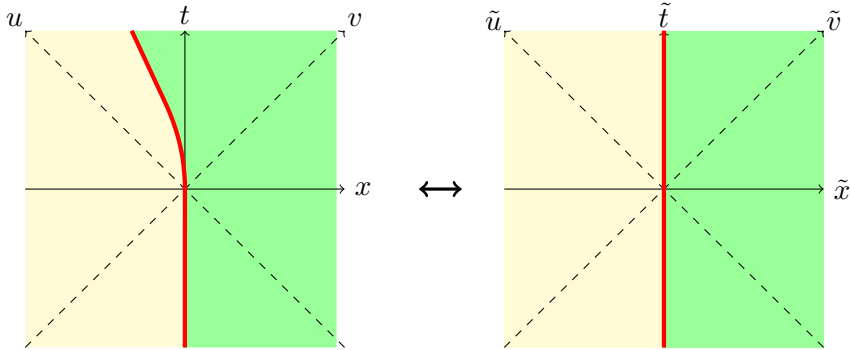
\begin{figure}[!h]
\centering
\begin{tikzpicture}[scale=0.7]
\fill[fill=green!40] (2.85,3)-- plot [smooth] coordinates {(-1, 3) (-0.90518, 2.7948) (-0.8419, 2.6580) (-0.74715, 2.4528) (-0.65233, 2.2476) (-0.55751, 2.0424) (-0.46269, 1.837) (-0.36787, 1.6321) (-0.27741, 1.4225) (-0.19658, 1.2034) (-0.1269, 0.9730) (-0.0703, 0.7296) (-0.02880, 0.47119) (-0.004837, 0.19516) (0,0) (0,-1) (0,-2) (0,-3)} --(2.85,-3);
\fill[fill=yellow!20,] (-3,3)--plot [smooth] coordinates {(-1, 3) (-0.90518, 2.7948) (-0.8419, 2.6580) (-0.74715, 2.4528) (-0.65233, 2.2476) (-0.55751, 2.0424) (-0.46269, 1.837) (-0.36787, 1.6321) (-0.27741, 1.4225) (-0.19658, 1.2034) (-0.1269, 0.9730) (-0.0703, 0.7296) (-0.02880, 0.47119) (-0.004837, 0.19516) (0,0) (0,-1) (0,-2) (0,-3)}--(-3,-3);
\draw [ ->] (0,-3) -- (0,3) node at (0,3.3) {$ t$};
\draw [ ->] (-3,0) -- (3,0) node[right] {$x$};
\draw [dashed, ->] (-3,-3) -- (3,3) node at (3.2,3.2) {$v$};
\draw [dashed, ->] (3,-3) -- (-3,3) node at (-3.2,3.2) {$u$};
\draw[<->, very thick] (4.4,0)--(5.2,0);
\draw[red,ultra thick] plot [smooth] coordinates {(-1, 3) (-0.90518, 2.7948) (-0.8419, 2.6580) (-0.74715, 2.4528) (-0.65233, 2.2476) (-0.55751, 2.0424) (-0.46269, 1.837) (-0.36787, 1.6321) (-0.27741, 1.4225) (-0.19658, 1.2034) (-0.1269, 0.9730) (-0.0703, 0.7296) (-0.02880, 0.47119) (-0.004837, 0.19516) (0,0) (0,-1) (0,-2) (0,-3)};
\fill[fill=green!40] (9,3)--(12,3)--(12,-3)--(9,-3);
\fill[fill=yellow!20,] (6,3)--(6,-3)--(9,-3)--(9,3);
\draw [ ->] (9,-3) -- (9,3) node at (9,3.3) {$\tilde{t}$};
\draw [ ->] (6,0) -- (12,0) node[right] {$\tilde{x}$};
\draw [dashed, ->] (6,-3) -- (12,3) node at (12.2,3.2) {$\tilde{v}$};
\draw [dashed, ->] (12,-3) -- (6,3) node at (5.8,3.2) {$\tilde{u}$};
\draw[red,ultra thick] (9,3) -- (9,-3);
\end{tikzpicture}
\caption{Conformal mapping for receding mirrors.}
\label{fig:receding}
\end{figure}

In all the previous computations, we took our bulk point at $(u,0)$ and then analyzed behaviour of OTOC's as a function of $u$. Here, instead of taking $v=0$ for the bulk point we will take $v=\frac{1}{2\pi T_H}\left(1-e^{-2\pi T_H L}\right)\equiv c_0$ which is the value of $p(L)$.

We perform the usual conformal transformation $\tilde{u}=p(u)$ and $\tilde{v}=v$ to map the dynamical mirror profile to a static one. Let us take the bulk point to be $P\equiv \left(\tilde{u},\tilde{v}\right)$, the reflected point would be $P_r\equiv (\tilde{v},\tilde{u})$. We take our boundary points to be $P_1\equiv \left(\tilde{v}_1,\tilde{v}_1\right)$ and $P_2\equiv \left(\tilde{v}_2,\tilde{v}_2\right)$. We will fuse the two boundary points at a same point $P_1\equiv P_2$. The cross-ratio becomes
\begin{equation}
\begin{split}
\eta&=i\tilde{\epsilon}_{12}\frac{\left(\tilde{u}-\tilde{v}\right)}{\left(\tilde{u}-\tilde{v}_1+i\tilde{\epsilon}_{01}\right)\left(\tilde{v}-\tilde{v}_1+i\tilde{\epsilon}_{02}\right)}\\
&=\frac{\big(p(u)-v\big)i\,\tilde{\epsilon}_{12}}{\big(p(u)-v_1+i\,\tilde{\epsilon}_{01}\big)\big(v-v_1+i\,\tilde{\epsilon}_{02})}\,.
\end{split}
\end{equation}

\subsection*{Case-1: $v=c_0, u=0$}
\begin{equation}
\begin{split}
\eta&=i\,\tilde{\epsilon}_{12}\frac{ -c_0}{(-v_1)(c_0-v_1)} \\
&=i\,\tilde{\epsilon}_{12}\frac{ c_0}{(v_1)(c_0-v_1)}\,.
\end{split}
\end{equation}
As in the previous cases, we take $v_1>0$ and $v_1<c_0$. 

\subsection*{Case-2: $v=c_0,~~ u=L-\alpha,~~ \alpha\rightarrow 0$}
In this case, we can't take $u\rightarrow \infty$ limit. Instead, we will take $u=L-\alpha$ where $\alpha$ is a small positive number and we will take $\alpha\rightarrow 0$.
\begin{equation}
\begin{split}
\eta&=i\tilde{\epsilon}_{12}\frac{\frac{1}{2\pi T_H}\left(1-e^{-2\pi T_H(L-\alpha)}\right)-c_0}{\left(\frac{1}{2\pi T_H}\left(1-e^{-2\pi T_H(L-\alpha)}\right)-v_1\right)\left(c_0-v_1\right)} \\
&=i\tilde{\epsilon}_{12}\frac{-\frac{1}{2\pi T_H}e^{-2\pi T_H L}2\pi T_H \alpha}{\left(c_0-v_1\right)\left(c_0-v_1\right)}\\
&=i\tilde{\epsilon}_{12} \frac{-e^{-2\pi T_H L}\alpha}{(c_0-v_1)^2}\,.
\end{split}
\end{equation}
So we see that $\eta$ approaches $0$ from the opposite directions for Case-1 and Case-2.

\subsection*{Case-3: $v=c_0, u=u_1=p^{-1}(v_1)$}
\begin{equation}
\begin{split}
\eta
&=i\tilde{\epsilon}_{12}\frac{p(u_1)-c_0}{[p(u_1)-v_1+i\tilde{\epsilon}_{01}][c_0-v_1+i\tilde{\epsilon}_{02}]}\\
&=-\frac{\tilde{\epsilon}_{12}}{\tilde{\epsilon}_{01}}=1+\frac{\tilde{\epsilon}_{20}}{\tilde{\epsilon}_{01}} \ .
\end{split}
\end{equation}

The cross ratio $\eta$ goes to zero from the opposite direction in the limit $u\rightarrow 0$ and $\alpha\rightarrow 0$. 
For the intermediate point $u=u_1$, in the out of time order case, $\tilde{\epsilon}_2>\tilde{\epsilon}_0>\tilde{\epsilon}_1$ or, $\tilde{\epsilon}_1>\tilde{\epsilon}_0>\tilde{\epsilon}_2$, it crosses the real axis after $\eta=1$ branch point.

For a large-$c$ CFT, as before, we can again use the Identity block result. We can very easily see that as $\alpha\rightarrow 0$:
\begin{equation}
\begin{split}
\mathcal{F}_0(\eta)
&=\left(\frac{1}{1-\frac{24\pi i h_w}{c\eta}}\right)^{2h_v}\\
&=\left(\frac{1}{1+\frac{24\pi  h_w (c_0-v_1)^2}{c\,\tilde{\epsilon}_{12}e^{-2\pi T_H L}\alpha}}\right)^{2h_v}\\
&\approx 1-\frac{48\pi h_v h_w (c_0-v_1)^2}{c\,\tilde{\epsilon}_{12}e^{-2\pi T_H L}}\left(\frac{1}{\alpha}\right)+\cdots\,.
\end{split}
\end{equation}
In this case, we again get polynomial behaviour.

\bibliographystyle{JHEP}
\bibliography{references}
\end{document}